%% file: paper-ngMEM.tex
\documentclass[fleqn,usenatbib,useAMS]{mnras}

\usepackage{epsfig,graphicx,natbib}
\usepackage{longtable}
\usepackage{amssymb}
\usepackage{amsmath}
\usepackage{color}
\usepackage{subcaption}
\usepackage{booktabs}

\usepackage{xspace}
\usepackage{float}
\usepackage{longtable}
\usepackage{breqn}
\usepackage{algpseudocode}
\usepackage{algorithm}
\usepackage{hyperref}
\usepackage{cases}
\usepackage{tabularx}
\usepackage[LGR,T1]{fontenc}
\usepackage{ae,aecompl}
\usepackage{xcolor}

\usepackage[normalem]{ulem} 

{\bf}{\it}
\newtheorem{definition}{Definition}{\bf}{\rm}
{\it}{\rm}
{\it}{\rm}
{\bf}{\it}
{\it}{\rm}
\newtheorem{problem}{Problem}{\it}{\rm}
{\bf}{\it}
{\it}{\rm}
\newtheorem{remark}{Remark}{\it}{\rm}
{\it}{\rm}

\newcommand{\RR}{\mathbb{R}}

\newcommand{\sgra}{Sgr~A$^*$\xspace}
\newcommand{\nxcorr}{\texttt{nxcorr }}
\newcommand{\nxcorrm}{\texttt{nxcorr}}

\newcommand{\fobs}{F_{\text{obs}}}
\newcommand{\fmod}{F_{\text{mod}}^{l}}
\newcommand{\fmods}{F_{\text{mod}}}
\newcommand{\vmod}{\mathcal{V}^{\text{mod}}}

\newcommand{\mfront}{\mathcal{K}-\text{front}}
\newcommand{\textgreek}[1]{\begingroup\fontencoding{LGR}\selectfont#1\endgroup}

\DeclareRobustCommand*{\citef}{\cite}

\title[Dynamic interferometric reconstruction with ngMEM]{New-generation Maximum Entropy Method (ngMEM): a Lagrangian-based algorithm for dynamic reconstruction of interferometric data}

\author[A.~Mus]{Alejandro~Mus$^{1,2}$
\thanks{Contact e-mail: 
\href{mailto:alejandro.mus@uv.es}{alejandro.mus@uv.es}}
and 
Ivan~Mart\'i-Vidal$^{1,2}$%
\\
$^{1}$ Departament d’Astronomia i Astrof\'isica, Universitat de Val\`encia, C. Dr. Moliner 50, 46100 Burjassot ,Val\`encia, Spain
\\
$^{2}$ Observatori Astron\`omic, Universitat de Val\`encia, Parc Cient\'ific, C. Catedr\'atico Jos\'e Beltr\'an 2, 46980 Paterna, Val\`encia, Spain
}

\date {Received  / Accepted}

\begin{document}
\label{firstpage}
\pagerange{\pageref{firstpage}--\pageref{lastpage}}
\maketitle

\begin{abstract}
Imaging interferometric data in radio astronomy requires the use of non-linear algorithms that rely on different assumptions on the source structure and may produce non-unique results. This is especially true for {Very Long Baseline Interferometry}  (VLBI) observations, where the sampling of Fourier space is very sparse. A basic tenet in standard VLBI imaging techniques is to assume that the observed source structure does not evolve during the observation. However, the recent VLBI results of the supermassive black hole (SMBH) at our Galactic Center (Sagittarius A$^*$, \sgra), recently reported by the Event Horizon Telescope Collaboration (EHTC), require the development of dynamic imaging algorithms, since it exhibits variability at minute timescales. In this paper, we introduce a new non-convex optimization problem that extends the standard Maximum Entropy Method (MEM), for reconstructing intra-observation dynamical images from interferometric data {that evolves in every integration time}. We present a rigorous mathematical formalism to solve the problem via the primal-dual approach. We build a Newton strategy and we give its {numerical} complexity. We also give a strategy to iteratively improve the obtained solution and finally, we define a novel figure of merit to evaluate the quality of the recovered solution. Then, we test the algorithm, called ngMEM, in different synthetic datasets, {with} increasing difficulty. Finally, we compare it with another well-established dynamical imaging method. {Within this comparison we identified a significant improvement of the ngMEM reconstructions. Moreover, the evaluation of the integration time evolution scheme and the time contribution showed to play a crucial role for obtaining good dynamic reconstructions.}
\end{abstract}


\section{Introduction}

{Very Long Baseline Interferometry} (VLBI) is a technique that {combines} telescopes, separated by large distances, to synthesize the aperture of a single telescope with size similar to the maximum distance among the VLBI elements~\citep{Thompson17}. With such large apertures, it is possible to obtain images of radio sources with resolutions as high as a few tens of $\mu$as {\citep[e.g.][]{EHTM871, EHT1, Gomez22, Lu23}}. However, VLBI only sparsely samples the Fourier components of an image, which may introduce severe degeneracies in the {ill-possed} inverse problem of image deconvolution \citep{Kellermann01,Ryle62,Thompson17}. In addition to this, Earth rotation synthesis assumes that the source being imaged is static during the experiment, typically hours. But there are cases of sources showing fast structural changes even in intra-hour observation, as it happens with the supermassive black hole (SMBH) Sagittarius A$^*$ (\sgra) in the Galactic Center (GC), recently imaged by the Event Horizon Telescope Collaboration (EHTC)~\citep[][]{EHT3, Knollmueller}. 
This source has an estimated mass of $M=4\times10^6 M_\odot$~\citep{Ghez08,Gillesen09}, and its gravitational timescale is only a few seconds. Moreover, its innermost stable circular orbit has periods of only 4 to 30 minutes, depending on the black hole spin~\citep{Bardeen72}. Even its polarization presents fast variations on similar timescales~\citep[e.g.][]{Gravity2023, Eckart06, Johnson15, Marrone06, Wielgus22, Zamaninasab10}.

Reconstructing images via deconvolution from sparse arrays, as the Event Horizon Telescope (EHT), is an ill-posed inverse problem based in some hypothesis on the underlying image, {for example, that the source is not evolving during the observation}. For static sources, the traditional techniques, as those based on CLEAN~\citep{Clark80, Hogbom1974, Mueller23a}, and those based on Maximum Entropy Methods (MEM)~\citep{mem,HoldawayWardle1988}, have proven to work very well for decades. However, if the source is changing in time, measurements are no longer satisfying the imaging {assumptions}, and these reconstruction algorithms may fail.

In the context of VLBI, there has been a growth on the development of different algorithms to image time-variable sources. Conventional snap-shot imaging methods have been used to reconstruct micro-quasars~\citep[for instance][]{MillerJones19} and many other stellar objects.\\
\cite{Lu16} presented a more complex strategy, a pioneer work on recovering a (2+1)-dimensional\footnote{The third dimension corresponds to time.} image based on a time-averaged morphology of \sgra in EHT scales. That is, they take into account the time dependence of the source during the imaging process. Recently, the~\cite{EHT3} applied a similar strategy to obtain a time-averaged image of this source during long time observations. However, these techniques present some limitations as the variability of the source injects a noise on the data hard to {model}.\\
To address this challenge, a new generation of imaging algorithms has emerged, aiming to overcome the rapid variability by reconstructing time-resolved brightness distributions, often referred to as ``movies''~\citep[e.g.][]{Arras22, starWarps, Farah22, Johnson17, Gao23,  Levis21, Levis22, Mueller23b}.

{In the near future, with denser arrays as the next-generation EHT}~\cite{Doeleman19, Johnson23a, Johnson23b}\footnote{See \url{https://www.ngeht.org/} for more information} (ngEHT), together with the increment of receiver sensitivity and computer capabilities, it will become more standard to reconstruct not only static images but movies of the sources. In this work, we present a new imaging algorithm,  {the new-generation Maximum Entropy Method (ngMEM), rooted in the MEM principle. It extends the optimization framework proposed by~\cite{mem} to encompass both static and dynamic reconstructions of evolving sources.}

{In addition, this paper surveys diverse methods for dynamic VLBI imaging, extensively discusses minimizing the ngMEM using a primal-dual concept, explores often-overlooked convergence criteria, strategies for initiating the minimization and ensuring its convergence.}

{The organization is as follows:} First, in Sect.~\ref{sec:vlbibases}, we recall the VLBI basics. Then, in Sect.~\ref{sec:static_imaging}, we present the {main} families for static imaging and, in more detail, the MEM appearing in~\cite{mem}. In Sect.~\ref{sec:stateart}, we show a brief overview of some of the actual dynamic imaging algorithms. In Sect.~\ref{sec:ngmem_formalism}, we present the ngMEM mathematical formulation. {We give a numerical method to solve it, we state its convergence and we discuss the initial point dependence. Then,} we compare it with current similar algorithms. In Sect.~\ref{sec:synthetic_data}, we apply the algorithm in two complex synthetic data cases: a ring with an orbiting hotspot of 24 hour period and a ring with a transient hotspot, emulating the behavior of \sgra inferred from the results presented in~\cite{Wielgus22}. To conclude the evaluation of the ngMEM, in Sect.~\ref{sec:mj_comparison} we conduct a fiducial benchmark by comparing its results with those obtained using a well-established method. Finally, Sect.~\ref{sec:summary} presents the conclusions and outlooks of this work.

\section{VLBI measurements and imaging}\label{sec:vlbibases}

An interferometer consists of $T$ telescopes observing the same source at the same time. Any pair of telescopes (which constitutes a two-element interferometer) is called baseline. Each baseline, at each integration time, measures a complex visibility, $\mathcal{V}_{ab}$, where $a$ and $b$ denote the telescope indices. Such visibilities are the time-averaged cross-correlation of the recorded scalar electric field at two telescopes, and they are related to the brightness distribution on the sky, $I\left(x\right)$, by the van Cittert-Zernike Theorem~\citep{Thompson17}
\begin{equation}
    \mathcal{V}\left(u,v\right):=\displaystyle{\int}I\left(x\right)e^{-2\pi i \left(xu+yv\right)}\ dx,
\end{equation}
where $\left(x,y\right)$ are the director cosines centered on the source, in radians, and $\left(u,v\right)$ the baseline vector, in wavelengths, projected {on the sky plane}.

In the case where the image presents structural time-dependent changes during the observation, the sampled visibilities{ $\mathcal{V}_t$} at a fixed time {$t$} only represent the corresponding instantaneous image or ``snapshot". Therefore, a set of snapshots can be reconstructed along the experiment. If $T$ is small, as in the case of the EHT in the 2017 {campaign}, where 7 and 8 sites {were} observing the SMBH Messier87* (M87*) and \sgra respectively~(see for instance~\citealt{EHTM871},~\citealt{EHT1}), there may be strong degeneracies that prevent a robust imaging of a time-variable source.

\section{Static imaging reconstruction}
\label{sec:static_imaging}

In this section, {we give a brief overview of the main methods used in static imaging reconstruction to contextualize and motivate dynamic imaging methods.}~{As in }~\cite{EHTM873}, we distinguish {two} main families of static imaging methods: {backward modeling and forward modeling.}

\subsection{Backward modeling}
CLEAN~\citep{Hogbom1974} and its variants~\citep[for instance][]{Clark80} are one of the most widely used algorithms for imaging. These iterative deconvolution methods are widely used for imaging and heavily rely on algorithmic parameters such as the CLEANing regions. In addition, they lack a closed mathematical formulation. Another significant limitation is their assumption that the source can be decomposed into a set of finite delta components. This may be insufficient to accurately describe extended image features observed in real astronomical images{~\citep{Arras21, Mueller23a}}.

There have been efforts to overcome this limitation by using multiscalar algorithms that model the image as a sum of extended basis functions of different scales~\citep{Bhatnagar04, Cornwell08, Rau11, Starck1994, Wakker1998}. Although these methods, called MS-CLEAN {for} multiscalar CLEAN, represent a significant advancement in imaging, they have not been extensively utilized in frontline VLBI. This is primarily due to the challenges associated with selecting appropriate basis functions, as different scales are sensitive to different regions of the uv-coverage. Additionally, MS-CLEAN methods do not inherently address the issue of missing regularization in CLEAN, specifically the occurrence of unphysical fits in the gaps of the uv-coverage. Recently~\cite{Mueller23a} developed a new MS-CLEAN algorithm to overcome this issue.

\subsection{Forward modeling}

In a first instance, we have focused on the non-convex optimization methods. These algorithms offer a high degree of flexibility by allowing the integration of complex data relations. Within this family, two main types can be identified: constrained optimization methods and unconstrained (penalty) methods, also called Regularized Maximum Likelihood (RML) methods.

\subsubsection{Non-linear constrained optimization}
On the one hand, constrained optimization methods, such as the Maximum Entropy Method~\citep[e.g.][]{mem, Gull85, Narayan86} and its variants, may include non-linear constraints through the use of Lagrange multipliers.\\
In general, they seek to maximize some metric, $S\left(p\right)$ subject to a set of constraints $g_i\left(x\right)$, and their typical formulation is
\begin{problem}
\label{prob:constrained_mem}
  \begin{equation*}
  \tag{Constrained}
    \begin{aligned}
      & \underset{p}{\text{maximize}}
      & & S\left(p\right),\\
      & \text{subject to}
      & & g_i\left(x\right)=0.
    \end{aligned}
  \end{equation*}
\end{problem}

\subsubsection{Non-linear unconstrained optimization}
On the other hand, RML methods can represent complex data relations {via regularizers} such as the $\ell_1$ norm~\citep{Honma14} or the total squared variation~\citep{Kuramochi18}, or even combinations~\citep[as][]{smilia, ehtima}. {These functionals may not be easily modeled as constraints}. The general formulation {of the RML methods} is
\begin{problem}
\label{eq:dual_mem_timeindp_other_version}
  \begin{equation*}
  \tag{RML}
    \begin{aligned}
      & \underset{p}{\text{minimize}}
      & & J := f - {\displaystyle \sum_k}\alpha_k \mathcal{R}_k\left(p\right),\\
    \end{aligned}
  \end{equation*}
\end{problem}
where $f$ is {the data fidelity term} (in general, the $\chi^2$ distribution based on visibilities, closure phases, closure amplitudes\footnote{Closures are quantities robust against calibration effects, see~\cite{Thompson17}}, etc), $\alpha_k$ are fixed penalty parameters (or regularizers) and $\mathcal{R}\left(p\right)$ are conditions (as the entropy, smoothness, sparsity) on the pixel distribution $p$.\\
A fine tuning on the penalty terms $\alpha_k$ is performed in order to find the best combination, which may depend on the particularities of the Fourier coverage and the source structure.\\
Recently,~\cite{MullerMus}~\cite{MusThesis}~\cite{Mus23} have presented a multiobjective optimization problem that decomposes Prob.~(\ref{eq:dual_mem_timeindp_other_version}) in $k$ optimization problems that are solved at the same time. In this way, {a} pararmeter survey to find the best values of $\alpha_k$ is not needed.

\subsubsection*{MEM: Cornwell version}
\label{subsec:mem_ind}

\cite{mem}{ presented a pioneering use of the Maximum Entropy in radio imaging.} In their work, {the MEM is formulated} in the form of Prob.~\eqref{prob:constrained_mem}.

If we define a set of visibilities, $\mathcal{V}:=\left\{\mathcal{V}_i\right\}$, with associated statistical weights, $\left\{\omega_i\right\}$ and we consider the vector of pixel intensities $p:=\left\{p_k\right\}$, located at the position offsets $\left\{(x_k, y_k)\right\}$ on the sky (with respect to the phase center of the correlation), we fix the objective function
\begin{equation}
J \colon \RR^{n}_{+} \to \RR_{+};\ p \mapsto S\left(p\right),
\end{equation}
{being} $S\left(p\right)$ a statistical measure.

A typical choice is the relative entropy~\citep[or the Kullback–Leibler, KL, divergence,][]{Kullback1951} $S\left(p\right):=-{\displaystyle \sum_k}p_k\log \dfrac{p_k}{M_k}$, where $M_k$ is the $k$-th pixel of the a priori model, $M$. Alternative versions for $S\left(p\right)$ and a deeper discussion can be found in~\cite{Narayan86, Thompson17}.

To adjust the entropy to the data, two equality constraints are established: the expected total flux, $\fmods$, should be equal to the recovered flux, $\fobs$, and the $\chi^2$ distribution of the observed visibilities $\mathcal{V}$ with respect to a model $\vmod$ should be close to its expected value $\Omega$. Recall that the $\chi^2$ of the visibilities is defined as

\begin{equation}\label{eq:chisq_expl_time_ind}
    \chi^2 := \sum_j{\omega_j \left| \mathcal{V}_j - \sum_k{p_k e^{\alpha_{jk}}}  \right|^2},~~\mathrm{with}~~\alpha_{jk} = \frac{2\pi} i(x_k u_j + y_k v_j).
\end{equation}
\noindent In this equation, $u_j$ and $v_j$ are the $\left(u,v\right)$ coordinates (in units of wavelength) of the $j$-th visibility, ($x_k, y_k$) are given in radians and $i$ is the imaginary unit.

With this setting,~\cite{mem} state the following non-convex constrained optimization primal problem:

\begin{problem}[Primal Problem]
\label{eq:mem_standard}
  \begin{equation*}
  \tag{MEM}
    \begin{aligned}
      & \underset{p}{\text{maximize}}
      & & -{\displaystyle \sum_k}p_k\log \dfrac{p_k}{M_k},\\
      & \text{subject to}
      & & \chi^{2}-\Omega=0,\\
      & & & \fmods- {\displaystyle \sum_k}p_k = 0.
    \end{aligned}
  \end{equation*}
\end{problem}

To solve Prob.~\eqref{eq:mem_standard}, the Lagrangian dual approach is used, whose formulation reads

\begin{problem}[Dual Problem]
\label{eq:dual_mem_timeindp}
  \begin{equation*}
  \tag{MEM-dual}
    \begin{aligned}
      & \underset{p}{\text{maximize }}\underset{\lambda,\beta}{\text{minimize}}
      & & L\left(\lambda,\beta,p\right) := -{\displaystyle \sum_k}p_k\log \dfrac{p_k}{M_k} - \\
      & & & \lambda\left(\chi^2-\Omega\right) - \beta\left(\fmods- {\displaystyle \sum_k}p_k\right),\\
    \end{aligned}
  \end{equation*}
\end{problem}

\noindent where $\lambda$ and $\beta$ are Lagrange multipliers.

\cite{mem} solve the optimization problem using a quasi-Newton method and updating the Lagrange multipliers in each iteration {such that} the Lagrangian is minimized for $\lambda$ and $\beta$, and thanks to this fact, the cost function is maximized with respect to the $p$ vector. Because of technical limitations, they work with a Hessian matrix with entries only in the superior, principal and inferior diagonal. For a sparse array like the EHT, which produces a poor dirty image, an important amount of information may be lost by doing the Hessian simplification. {Our work extends this problem}.

Note that this is a non-convex constrained optimization problem, since the $\chi^2$ constraint is not affine~\citep{boyd04}. This may produce, among other consequences, a set of local minima instead of a unique global minimum. Therefore, special caution must be taken during the process of solving it, in particular using \textit{local search methods}, i.e., algorithms that use the information of the gradient or the Hessian. For those cases, a starting point near a locally convex neighborhood is needed to guarantee the convergence to a local minimum. In interferometric observations at optical wavelengths, image reconstruction algorithms that overcome this limitation (by recovering a set of local minima) are already available~\citep[e.g][]{Ozon16}. More recent results have also been published for VLBI~\citep{MullerMus}.

Extensions of the formulation by \cite{mem} {have not only been used} in the radio regime. In~\cite{Buscher1994}, {the} authors developed the MEM in the optical regime, to take into account the bispectrum, which later inspired the VLBI imaging community the inclusion of closure quantities and MEM-based priors~\citep{smilia, ehtimc}.

\cite{HoldawayWardle1988} and~\cite{HoldawayPhD} extended for the first time the method proposed by~\cite{mem} to include the polarized entropy defined in~\cite{Gull85} for the reconstruction of polarization images from VLBI data. More recent works have been done in this area, for example~, ~\cite{smilib},~\cite{ehtima}
~\cite{Coughlan13} and~\cite{Mus23}. Polarimetry is out of the scope of this paper, and we refer the interested reader to the bibliography.

\subsection{Bayesian and machine learning algorithms}
The third family of interferometric deconvolution algorithms {is based} on Bayesian exploration of the posterior distribution of the source's parameters structure, using techniques as Markov Chain Monte Carlo (MCMC). They aim to find the reconstruction with the best representation of the signal and have the capability of providing uncertainty quantification. Some recent examples are~\cite{Arras21},~\cite{ThemisRef} and~\cite{Tiede22}.

Further works have been done in forward-modeling algorithms~\citep[e.g.][]{Aghabiglou23,Gao23,Terris2023} but it is out of the scope of this paper to give a full reference of them.

\section{Some previous works on dynamic imaging}\label{sec:stateart}

The development of novel dynamic imaging algorithms has been active in the area driven by the EHT science. The possibility of observing \sgra with the resolution of the EHT, has been a strong motivation for the development of such algorithms.{~\cite{Roelofs23} designed a challenge of synthetic datasets based on realistic observations and performed benchmark comparison of different static and dynamic reconstructions}.

In this section, we give a brief overview of previous works on dynamic imaging following the classification of Sect.~\ref{sec:static_imaging}. It is out of the scope of this paper to give a full review of all the methods. Instead, the presented summary aims to contextualize the algorithm we introduce in this work.

\subsection{Backward modeling}

Recently, there has been an effort to extend the traditional CLEAN algorithm to a dynamic scale, for being able to reconstruct movies. The common technique implies that, after the imaging process reaches convergence, based on a specified stopping criterion, and produces an average static image, the CLEAN dynamic imaging technique is employed. This involves dividing the observation into smaller portions, referred to as ``snapshots'', with a time duration similar to the expected timescale variability of the sources. Assuming minimal structural changes over time, the model corresponding to the static image serves as the initial model. The goal is to identify structural changes by cleaning the residual map associated with each data snapshot~\citep[for an example of a similar method, see][]{Mus22, Stewart2011}. However, this technique may be very limited in the cases of the sparse uv-coverage, which is the case of the EHT.~\cite{Farah22} introduce a new metric to overcome this limitation that ranks array configuration by their ability to produce accurate CLEAN dynamical reconstructions.

\subsection{Forward modeling}

In the context of non-linear optimization methods, efforts have been devoted to extend the modelization of the problem to take into account the kinematics of the source. These efforts have been mainly done for the RML methods. Their flexibility allows to add new regularizers (or penalty terms) into the cost functional to deal with kinematics of the source~\citep[see for example][]{Johnson17}. In these methods, the observation can be thought as a set of static snapshots (or frames) that compose the ``movie''. These methods usually define a metric over the set of frames, and the regularizer terms impose similarity between adjacent frames and penalize the distant {ones}. The inclusion of temporal regularization in a model can result in the suppression of inherent source variability, but the extent to which this occurs depends on the specific source structure and timescale of variability. In Sect.~\ref{sec:mj_comparison}, we give two examples of time regularizers appearing in~\cite{Johnson17}. We refer to that section for more details.
Determining the threshold at which the regularization becomes excessive is subjective.

{A slightly different approach to model the optimization problem is presented} in~\cite{starWarps}, where they present \texttt{StarWarps}. This method utilizes a probabilistic graphical model to generate snapshots of a movie by solving the posterior probability distribution. This distribution is defined as a combination of three factors: data likelihood, multivariate Gaussian distributions for each snapshot, and transitional probabilities between adjacent snapshots. These transitional probabilities effectively serve as spatial and temporal regularization. By leveraging a Gaussian approximation, \texttt{StarWarps} allows for the exact inference of the movie by computing the mean and covariance of the image in nearby pixels in brightness and in time. In contrast, the RML dynamic reconstruction only derives a maximum a-posteriori estimation.

Other RML-based techniques have been developed by \cite{Mueller22a} (RML approach based on compressed sensing),~\cite{Mus23} (multiobjective optimization formulation extending the one presented in~\cite{MullerMus} including dynamic and polarimetric terms) and~\citep{Gao23} (neural networks). In all cases, all dynamic non-linear optimization methods solve instances of Prob.~\eqref{eq:dual_mem_timeindp_other_version}. In this work, we present a novel algorithm that deals with the kinematics by generalizing Prob.~\eqref{eq:mem_standard}.

{In the context of Bayesian algorithms,}~\cite{Arras22} extended the algorithm they presented in~\cite{Arras21}, \texttt{resolve}, to infer the time-variability of the source's brightness distribution by exploiting its correlation structure in time, and hence, assuming correlation in close-by frames. {Using this algorithm,~\cite{Knollmueller} recently {presented} first dynamic reconstructions of \sgra}.

\section{\MakeLowercase{ng}MEM formalism}
\label{sec:ngmem_formalism}

In this section, we present the mathematical formalism for our novel algorithm for the dynamic reconstruction of VLBI interferometric data, the so-called \textit{new-generation Maximum Entropy Method} (or ngMEM).  First, we introduce the main idea of the problem.
Then, we develop the mathematical formalism of the ngMEM, by extending the cost functional to take into account the kinematics of the source. Finally, we give the main differences between the current dynamical deconvolution algorithms and ours.

The algorithm we propose, as a generalization of Prob.~\eqref{eq:mem_standard}, addresses a non-convex optimization problem. The inclusion of the $\chi^2$ equality constraint results in the loss of convexity in the problem formulation, as it happens in~\cite{mem}.
Non-convexity is a characteristic that makes it especially difficult to determine and obtain, if it exists, the global optimal value of the cost functional. However, we can still benefit from the differentiability of the objective function and the constraints. It is beyond the scope of this paper to go into detail on the optimality conditions for non-linear optimization problems and to prove constraint qualifications~\citep[authors refer to][for a detailed discussion]{{bonnans06}}. Suffice it to say that our problem does satisfy sufficient and necessary optimality conditions, i.e., every local solution
has its paired Lagrange multipliers. . 
Therefore, if we find a solution on the dual problem, we have its paired primal solution.

\subsection{Time-dependent (dynamic) ngMEM}
\label{sec:dynamic_ngmem}

To extend the traditional MEM to reconstruct movies, we have included a new term in the cost function that constraints the evolution on the source brightness distribution penalized by a ``time regularizer''.
When the regularizer is zero, the method might produce an unrealistic variability (since we would be applying the MEM to each snapshot, independently). However, if the regularizer is too large, we would recover a quasi-static movie (since all pixels would be forced to remain as constant as possible). Later in this section, we explain in detail how we handle this new term.

Before presenting the functional to be optimized, let us divide the set of visibilities into $r$ subsets, defined as those where the visibilities were observed within a given time window. In particular, if $t$ is the observing time, 
\begin{equation}\label{eq:keyframes}
\mathcal{V}^l = \left\{\mathcal{V} ~\mathrm{for}~~ t \in [t_l - \Delta t_l/2, t_l + \Delta t_l/2] \right\}, ~~ l \in [1,r].
\end{equation}

In this equation, $\Delta t_l$ is a frame-dependent scalar that determines its duration and $\mathcal{V}^l$ are the visibilites belonging to such a frame.
Let us model these data as a set of ``image keyframes'', where there is one image keyframe for each data frame. The model will have a total of $r \times N^2$ parameters (i.e., $r$ images of $N^2$ pixels each). We can solve this problem of dynamic imaging (i.e., find the set of image keyframes that optimally fit the data) by using the following formalism. If $p_i^k$ is the $i$-th pixel of the $k$-th image keyframe, we can arrange all the model parameters into a ``supervector'' $\vec{Q}$, defined as

\begin{equation}
    \vec{Q} = \left\{ p_1^1, ~p_2^1, \cdots, ~p_{N^2}^1, ~p_1^2, ~p_2^2, \cdots, ~p_{N^2}^2, ~p_1^3, ~p_2^3, \cdots p_{N^2}^r    \right\}.
\end{equation}

In essence, we have ordered the pixels keyframe by keyframe into $\vec{Q}$.

\noindent With this, we can define what we call the Shannon~\citep{Shannon1949} ``time entropy'', $\mathcal{T}\colon \RR^{n}_{+}\times\left(\mathcal{V}^l\times \mathcal{V}^m\right)\to\RR_{+}$, which promotes the most constant evolution of the pixel brightness distribution during the whole experiment via

\begin{multline}
\label{eq:image_memory}
    \mathcal{T}\left(p\right) := \sum_{n, l \neq m}{T_n^{lm}},~~~\mathrm{where}~~~ \\
    T_n^{lm} := 
    e^{-\frac{|t^l - t^m|^2}{2\tau^2}}
    \left(|p_n^l - p_n^m |+C\right)\log{\left( |p_n^l - p_n^m| + C\right)},
\end{multline}

\noindent where $\tau$, which we call ``image memory'', will be discussed later and $C$ is a small positive constant to ensure finite values of the logarithm.
We notice that, for clarity, the super-indices refer to time keyframes, while the sub-indices refer to the pixel index in the image.
On the one hand, $\mathcal{T}$ preserves the same properties\footnote{continuity, differentiability and convexity} as $J$ with respect to the $p$ component. This is a great advantage.
Since the maximization of the original entropy (Prob.~\eqref{eq:mem_standard}) forces the image to have the minimum contrast that is compatible with the data \citep[i.e., a higher entropy corresponds to a smaller difference in the intensities among pixels, e.g.][]{mem}, adding the time entropy to the functional will similarly enforce the pixels to have a \textit{minimum time variation that is still compatible with the data}. \\
On the other hand, the exponential injects correlations between the intensity values that any given pixel may take at different times. For keyframes with time separations of the order of (or shorter than) $\tau$, the time entropy will force any pixel to have an intensity with a minimum time variation. However, for keyframes separated by a time lag longer than $\tau$, the time differences of the pixels will have little impact in the time entropy. Increasing the value of $\tau$ implies allowing the time entropy to force smaller time variability of the pixel intensities at longer timescales. {Gaussian weighting is just one possible way to leverage closer keyframe, but it has some advantages: 1) it is conceptually simple, 2) it is $\mathcal{C}^\infty$ (infinitely differentiable), and easy to be differentiated 3) it is not computationally expensive and 4) Gaussian distribution maximizes the entropy for a given mean and $\tau$ (therefore, we maximize entropy not only in the pixel time-dependent domain, but also in the frame distribution time-domain).}

Thus, we can construct the cost function as the sum of the two Shannon entropies
\begin{align}
J\  \colon \RR^{n}_{+}\times \mathcal{V}^l &\to \RR_{+} \\
\left(p,t\right) &\mapsto {\displaystyle \sum_{k,l}}p^{l}_{k}\log p^{l}_{k} + \mu\mathcal{T}\left(p\right),
\end{align}

with $\mu\in\RR_{+}$.

Henceforth, the optimization problem now reads

\begin{problem}[ngMEM Problem time-dependent]
\label{eq:mem_expl_time_dep}
  \begin{equation*}
  \tag{ngMEM}
    \begin{aligned}
      & \underset{p}{\text{minimize}}
      & & J\left(p\right) := {\displaystyle \sum_{k,l}}p^{l}_{k}\log p^{l}_{k} + \mu\mathcal{T}\left(p\right),\\
      & \text{subject to}
      & & \chi^{2} - \Omega = 0,\\
      & & & \fmod- {\displaystyle \sum_k}p^{l}_k = 0,\quad \forall \text{ keyframe } l.
    \end{aligned}
  \end{equation*}
\end{problem}

\noindent where $\fmod$ is the total flux density of the source at keyframe $l$ and the parameter $\mu$ is the ``weight'' of the time entropy, meaning that larger values will force a smaller time variability in the model. This number is fixed before the optimization problem (it is a constant for $J$). \\
The source model is computed at each integration time (or ``frame'', which shall not be confused with the ``keyframes'' in Eq~\eqref{eq:keyframes}), by using a pixel-wise time interpolation of the keyframes. For instance, if we use a linear time interpolation, the model image at time $t$, with $t \in (t^l, t^{l+1})$, is

\begin{equation}
p_k(t) = \frac{1}{t^{l+1}-t^l}( p_k^{l+1}(t-t^l) + p_k^l(t^{l+1}-t) ).
\end{equation}

This way, the $\chi^2$ is computed using the image interpolation at each exact visibility observing time (i.e., at each frame), even though one keyframe may include several integration times. 

The final reconstruction will be affected by how the keyframes are interpolated into each {integration time} (frame). Currently, two interpolation modes are implemented (nearest and piece-wise linear), although the algorithm can be easily adapted to other interpolation modes, like splines.

Moreover, we have a set of equality constraints on the flux density, which means that ngMEM is able to use lightcurve information as a constraint to the model. This feature may be {especially} important for EHT observations where ALMA participates, since precise lightcurves can be derived from the ALMA-only observations~\citep{Mus22, Wielgus22LC}. 

\subsection{Numerical method for solving the ngMEM problem}
It can be shown that the Lagrangian dual for Prob.~\eqref{eq:mem_expl_time_dep} is

\begin{problem}[Dual of the ngMEM Problem time-dependent]
  \label{eq:dual_mem_expl_time_dep}
  \begin{equation*}
    \begin{aligned}
      & \underset{p}{\text{minimize }}\underset{\lambda,\vec{\beta}}{\text{maximize}}
      & & L\left(\lambda,\vec{\beta},p,t\right) := {\displaystyle \sum_k}p^{l}_{k}\log p^{l}_{k} +\\
      & & &\mu\mathcal{T}\left(p\right) - \lambda\left(\chi^2-\Omega\right) - \\ 
      & & & {\displaystyle \sum_l}\beta^l\left(\fmod- {\displaystyle \sum_k}p^{l}_{k}\right).
    \end{aligned}
  \end{equation*}
\end{problem}

Details on the derivation of the dual problem and zero-gap proof are available in~\cite{MusThesis}. That means, the solution to Prob.~\eqref{eq:mem_expl_time_dep} can be recovered by solving Prob.~\eqref{eq:dual_mem_expl_time_dep}. We have denoted by $\vec{\beta}$ the vector of Lagrange multipliers acting on the lightcurve constraints and by $\beta^l$ the specific multiplier acting at keyframe $l$.

We solve Prob.~\eqref{eq:dual_mem_expl_time_dep} via Newton's method on a locally convex neighborhood. Thus, we need to compute the gradient and the Hessian of the Lagrangian. Since we already know how the entropy and $\chi^2$ derivatives look like~\citep[for instance][]{mem, Narayan86} {we just present the ones here} corresponding to the new term, $\mathcal{T}$.

The gradient is

\begin{equation}
    \frac{\partial \mathcal{T}}{\partial p_n^l} = \sum_{m \neq l}{e^{-\frac{|t^l - t^m|^2}{2\tau^2}}\left(1 + \log{(|p_n^l-p_n^m| + C)} \right)},
\end{equation}

\noindent and the second derivative is

\begin{equation}
    \frac{\partial^2 \mathcal{T}}{\partial p_n^l \partial p_u^m} = \begin{cases}
    0 ~~\mathrm{if}~~ l=m, \\
    \dfrac{\delta_{nu}}{{|p_n^j - p_u^k| + C}}e^{-\frac{|t^l - t^m|^2}{2\tau^2}} ~~\mathrm{otherwise.}
    \end{cases}
    \label{Teq}
\end{equation}

Notice that, according to the Kronecker delta, the second derivative of the time entropy is non-zero only if $n=u$ (i.e., the time entropy only relates the change in time of each pixel, but not among different pixels). The time entropy introduces non-zero terms in $\mathcal{H}$, which fall out of the diagonal boxes that are related to $\chi^2_l$. Actually, $\mathcal{H}$ can be written as a matrix of matrices, in the following way:

\begin{equation}
    \mathcal{H} = \begin{bmatrix}
    H^1 & T^{12}    & \cdots & T^{1r} \\
    T^{21} & H^2    & \cdots & T^{2r} \\
    \vdots & \vdots & \ddots & \vdots \\
    T^{r1} & T^{r2} & \cdots & H^{r} \\
    \end{bmatrix},
    \label{eq: hessian_time_dep}
\end{equation}

\noindent where $H^l,\ l=1,\ldots,r$ is the Hessian of the Lagrangian in the $l$-th image keyframe. 
Regarding the matrices $T^{jk}$, they can be computed from Eq.~\eqref{Teq}, namely:

\begin{equation}
    T^{jk}_{mn} = e^{-\frac{|t^j - t^k|^2}{2\tau^2}}\frac{\delta_{mn}}{|p_m^j - p_n^k| + C}. \label{eq: tijk}
\end{equation}

In a similar fashion to $\vec{Q}$, we create the ``super-gradient'', $\vec{K}$, by arranging the gradient of the Lagrangian, with respect to all the model parameters, 

\begin{equation*}
    K_m = \frac{\partial E_l}{\partial p^l_n} +\mu \frac{\partial \mathcal{T}}{\partial p^l_n} -  \lambda \frac{\partial \chi^2_l}{\partial p^l_n} -\vec{\beta} ~~~\mathrm{with}~~~ m = (l-1)\times N + n
\end{equation*}

\noindent being $\chi^2_l$ {and } $E_l$ the $\chi^2$ { and the KL divergence respectively in the}  $l$-th keyframe.

With all these definitions, the time-dependent equation for Newton becomes, for a given iteration $k$,

\begin{equation}
    \mathcal{H}_{k} \left(\vec{Q}_{k+1} - \vec{Q}_{k} \right)= -\vec{K}_{k},
\end{equation}

\noindent or in a more compact expression by setting $d_k:=\vec{Q}_{k+1} - \vec{Q}_{k}$,
\begin{equation}
\label{eq:newton_step}
    \mathcal{H}_{k}d_k= -\vec{K}_{k}.
\end{equation}

The set of equations in Eq.~\eqref{eq:newton_step} are the first optimal (Karush-Kuhn-Tucker) conditions {which ensures the existence of an optimal solution associated} to the so-called \textit{osculating quadratic problem} of Prob.~\eqref{eq:dual_mem_expl_time_dep} at $\left(p_k,\lambda_k,\beta_k\right)$~\citep[see][for more details]{bonnans06}

\begin{numcases}{}
      & $\underset{d}{\text{min }} \nabla_{p}L^{T}_{k}\left(\lambda_k,\vec{\beta}_k,p_k,t\right) d + \dfrac{1}{2}d^T \nabla_{p,p}L_k\left(\lambda_k,\vec{\beta}_k,p_k,t\right) d$\label{eq:osculating_quadratic_1} \\
      & $c_k + \nabla_{p}c_{k}d = 0,$\label{eq:osculating_quadratic_2}
\end{numcases}

where $c_k$ denotes the vector of the equality constraints.

The above system of equations generates a sequence of solutions $\left(d^{p}_{k},d^{\lambda}_{k},d^{\vec{\beta}}_{k}\right)$ to define the next iterate $\left(p_{k+1},\lambda_{k+1},\vec{\beta}_{k+1}\right)$. Methods that solve simultaneously primal and dual problems are called \textit{primal-dual methods}.
Of course, a solution of Eq.~\eqref{eq:newton_step} is dependent on the election of several factors as the starting iterate, $\mu$ and in a lesser way, $\tau$.

{In this framework, the ngMEM algorithm is as follows}

\begin{algorithm}
\caption{Movie reconstruction algorithm solving the ngMEM.}\label{alg:ngmem}
\begin{algorithmic}
\Require $p_0,\lambda_0,\vec{\beta}_0,\mu,\texttt{times},\texttt{npix},\texttt{niter}$
\State ${\texttt{keyFrames}} \gets \texttt{divideObservation(times);}$
\State $\tilde{\Omega} \gets \texttt{estimateNoise();}$
\State $F^{\texttt{obs}} \gets \texttt{getLightcurves();}$
\While{SSC not satisfied or $k=0,\ldots,$niter-1}
\State $p_{k+1},\lambda_{k+1},\vec{\beta}_{k+1} \gets \texttt{Solve Prob~\eqref{eq:dual_mem_expl_time_dep}}\left(p_{k},\lambda_{k},\vec{\beta}_{k},\texttt{times},\texttt{npix}\right)$
\EndWhile
\State $p^*,\lambda^*,\vec{\beta}^* \gets p_{k},\lambda_{k},\vec{\beta}_{k}$
\end{algorithmic}
\end{algorithm}

\noindent In the following paragraphs, we describe Alg.~\ref{alg:ngmem} in detail. {First, we take a look in a simple case}. Then, we present its inputs, its critical steps, etc. Finally, we explore and compare solutions for the free parameter $\mu$.

\subsection{A particular case: ngMEM with $\mu=0$}
\label{subsec:ngmem_mu_0}

{It is instructive to have a closer look at a special case $\mu=0$. Our optimization problem is an extension of the problem defined in~\cite{mem}, or Prob.~\eqref{eq:mem_standard}, but using a flat prior, the Shannon entropy. Indeed, if we take the model $M$ appearing in their formulation to be constant, the ngMEM could be seen as a perturbation of the form $\phi\left(p,\mu\right)$ in such a way that }
\begin{equation*}
    -\phi\left(p,0\right)=-{\displaystyle \sum_k}p_k\log p_k.
\end{equation*}

{It is noteworthy that the presented case is not equivalent to the standard snap-shot MEM imaging. The computation of the $\chi^2$ involves interpolation between adjacent keyframes. Consequently, even when performing the deconvolution without the time entropy regularizer, there is still a small amount of information interchange among neighboring keyframes.
}

{In Appendix~\ref{sec:time_dependent_ngmem}, we present the ``super-Hessian'' for this particular case and we compare a standard RML method and the ngMEM applied to EHT real observation taken during the 2017 campaign.}

\subsection{Algorithm considerations}\label{subsec:alg_considerations}

The inputs of the algorithm are a set of visibilites correspondent to an observation, the hyper-parameter $\mu$, initial values for the Lagrange multipliers, $\lambda_0,\vec{\beta}_0$, the number of pixels for each image, \texttt{npix}, \texttt{niter} or total number of iterations used as a stopping criterion control and the observation times, \texttt{times}.

\subsubsection*{Global philosophy and convergence}
The global philosophy is as follows: we first compute the estimated noise value for the $\chi^2$, $\tilde{\Omega}$, and we determine the keyframes (in Sect.~\ref{sec:synthetic_data} we explain the way we have used to estimate it and how we can create the keyframes, in more detail) and, for each of the keyframes, we compute the flux density centered in such a keyframe (for instance, from the intra-ALMA visibilities in ALMA-EHT observations), $\fmod$.\\
Second, we construct the Lagrange dual problem associated to Prob.~\eqref{eq:mem_expl_time_dep}.\\
Third, since objective and constraints are, at least, twice continuous and differentiable in their domain, we can find, for example, using a (quasi-) Newton method a value for $\lambda^*$ and $\vec{\beta}^*$ such that
$L\left(\lambda^*,\vec{\beta}^*,p^*,t\right)$ is minimized for some vector $p^*$ associated to the vector of multipliers $\left(\lambda^*,\vec{\beta}^*\right)$ by constructing a convergent sequence $\left\{\left(p_k,\lambda_k,\vec{\beta}_k\right)\right\}$ that deals simultaneously with the objective minimization and with the constraint satisfaction. We construct such sequence by solving the osculating quadratic problem (Eq.~\eqref{eq:osculating_quadratic_1} and Eq.~\eqref{eq:osculating_quadratic_2}). Quasi-Newton algorithms have the advantage to converge quadratically~\citep[see, for instance][]{bonnans06} in a convex neighborhood.\\
Finally, the Second order Sufficient Conditions (SSC) on non-linear problems \citep{bonnans06}, ensure that the vector $p^*$ paired with $\left(\lambda^*,\vec{\beta}^*\right)$ is a \textit{local} solution in a certain neighborhood of the optimization problem, Prob.~\eqref{eq:mem_expl_time_dep}. In case the sequence has not reached a point where the SSC are not satisfied, \texttt{niter} acts as stopping criterion.

\subsubsection*{Solving the osculating problem}

{For solving Eqs.~\eqref{eq:osculating_quadratic_1},~\eqref{eq:osculating_quadratic_2}} {in each iteration until convergence, we have used the Cholesky decomposition, which requires the Hessian to be positive definite. This is not a problem in our case, since the Lagrangian is convex with respect to $p$. Other strategies for solving the system could have stronger requisites (for example, to be diagonal dominant) and the value of $\mu$ - which may affect the value of the Hessian elements far from the diagonal - could be limited by such a requisite.}

\subsubsection*{Globalization}
However, the Prob.~\eqref{eq:mem_expl_time_dep} is non-convex. As discussed before, this implies that gradient-based strategies do not guarantee convergence to a global minimum. This property imposes, among others, the condition of finding a good starting point: (Quasi-) Newton methods {exploit} the geometry of the function (gradient and Hessian). Such algorithms generate converging sequences (to an optimal primal-dual solution) if the initial point is close enough to a stationary point of the Lagrangian; otherwise, they may get stuck in a neighborhood of a saddle point, or even they may not converge at all. For this, it is important to have techniques to force convergence, even when the first iterate is far from a solution. This process is called \textit{globalization}. We have implemented a globalization technique to control the $\chi^2$.

\subsubsection*{Starting point}
In this case, the dirty image is a natural choice for the initial point. Or even the result of the MEM associated to Prob.~\eqref{eq:mem_standard}. A common strategy in penalty methods is to iteratively compute the solution to the problem using the previous solution as a starting point and increasing the values of the penalty multiplier in each iteration. This method allows an increasing accuracy in the solution. Nevertheless,~\cite{Polyak71} computes a rate of accuracy and convergence of this kind of strategy and shows that they are not ever-increasing in accuracy: there is a limit where the gained accuracy is not worth it, compared to the computational cost used to find it.\\
Since $\mu$ can be seen as a penalty term in Prob.~\eqref{eq:mem_standard} and, thus, we are in a similar setting as in~\cite{Polyak71}, we have followed this iterative strategy to compute our initial point $p_0$. We start with the dirty image and we solve the ngMEM for $\mu=0$. The obtained solution is used as a starting point for the ngMEM with $\mu> 0$. Note that this process could be repeated iteratively, but we have found good solutions with just doing one iteration. In Fig.~\ref{fig:comparison_initialpoints} of the Appendix~\ref{sec:comparison_initial_points}, we show a comparison on the obtained solutions for the ngMEM starting with different initial points (the dirty image and the result of our iterative strategy) in one of the synthetic data cases presented in this work (see Sect.~\ref{sec:synthetic_data} for further details).\\
It is important to stress that, regardless of the starting point, if a local optimal $\left(\lambda^*,\beta^*\right)$ that minimizes the Lagrangian with respect to the multipliers is found by the algorithm, the associated primal solution $p^*$ is a valid movie reconstruction. However, it could be a local solution with a big room to be improved (in terms of the value for the cost function).

\subsection{ngMEM algorithm details}

{ngMEM depends on a number of free parameters and considerations, among them the selection of the keyframes and the relative weight of the time entropy, $\mu$, stands out. In the next paragraphs we discuss some algorithmic consideration regarding the selection of the keyframes and free parameters.}

\subsubsection*{Keyframe distribution}
Following Alg.~\ref{alg:ngmem}, we first need to split the experiment in keyframes. Our algorithm allows to either uniformly distribute the keyframes during the whole set of observations or imposing the keyframes in particular observing times. {The first option is the simpler approach}. It has some limitations, like creating ``jumps'' between different keyframes and not ensuring a linear and smooth transition of the times. The flexibility to select particular observing times for creating a keyframe allows, for example, to create keyframes using the scans as reference ({for instance, keyframes are positioned either in the middle of the scans or at the beginning and end of specific scans)}. Moreover, if the source presents kinematics during a particular interval of the observation, this second option allows to concentrate more keyframes in such period to recover a better evolution of the source~{\citep[for example][gave specific time windows that proved to be more effective for dynamic imaging reconstruction of \sgra during the 2017 EHT]{Farah22}}.

In general, not all keyframes will have the same number of visibilites, being some keyframes quite limited.

\subsubsection*{Free parameters}
The free parameters, $\mu$ and $\tau$, are a measurement for weighting the kinematics of the source during the reconstruction. In essence, they determine the cost associated of getting the most constant solution which still fits into the data.\\
Note that $\mu$ affects the whole time-dependent term of the cost function, while $\tau$ is a measure of the timescale of the correlation between the values that any given pixel takes at different times. For example, larger values on $\mu$ and $\tau$ in a short observation time will map into smaller time variability. {$\tau$ can be computed based on observational data or in theoretical results, as the innermost stable circular orbit (or ISCO) in the case of a black hole.} \\
The parameter $C$ regulates the extent to which the brightness of a pixel $p_n$ can vary between two keyframes $l$ and $m$. When the brightness of a pixel remains relatively constant, the corresponding time factor entropy, given by $\log\left(\lvert p^{l}_{n}-p^{m}_{n}\rvert\right)$, tends to infinity. Therefore, $C$ is used to prevent the emergence of meaningless values by constraining the difference between pixel intensities.

The second parameter we need to carefully find is the estimation of the expected noise level (the expected value for the $\chi^2$ which we call $\Omega$ in Eqs.~\eqref{eq:mem_expl_time_dep},~\eqref{eq:dual_mem_expl_time_dep}). A too low value makes the algorithm adjust noise, while a too large value may leave a fraction of the true source signal as part of the residuals.

The way we have designed to find the best $\mu$, number of keyframes and their location is by running a gridsearch. We choose a prior parameter space with a range for each of the parameters. In some cases, it will be possible to reduce the dimension of this space. Ideally, we want to reduce the parameter space as much as possible, since the algorithm will be called for each combination of free parameters.

\subsubsection*{Non-negativity intrinsic constraint}
Finally, there is a last intrinsic constraint: we impose that, for all keyframes $l$, in any iteration of the algorithm, all the components of the brightness vector $p^{l}_{k}$, have to be greater than a cutoff value, called \textit{minimum flux}. There are several options to fix this constant (for example, setting it as the rms). Nevertheless, having sparse arrays like the EHT, or long baselines, increases significantly this statistic. If, in addition, the source intensity is low, then we could be limiting ourselves. Therefore, depending on the case, this constant can be more or less conservative.

\subsection{Differences to previous strategies}

Our algorithm is philosophically and mathematically different from the RML and multiobjective optimization methods, since we use the dual approach. We present {major} points of difference between the ngMEM and the rest of algorithms.
\begin{itemize}
    \item {ngMEM is the first MEM-based algorithm including a time-entropy. As shown in Sec.~\ref{subsec:ngmem_mu_0}, it is an extension to the time domain of the traditional maximum entropy problem.} 
    \item Previous (RML and non-RML) methods involving the entropy, as the ones cited in Sect.~\ref{sec:static_imaging} and Sect.~\ref{sec:stateart}, work with the Kullback–Leibler divergence of the pixel brightness, i.e., they use the relative entropy with respect to a model. On the other hand, our algorithm computes it respect to a flat prior. Therefore, any contrast in the image and/or any time change in the pixel intensity, contributes to the total entropy, increasing the value of the cost function. We consider this approach to be a conservative one, in the sense that it minimizes the model time evolution that is still compatible with the data. 
    \item {ngMEM and RML differ in philosophy: ngMEM minimizes the entropy among all solutions that fit the data (direct implementation of Occam's razor) while RML methods try to balance data and regularization terms}.
    \item Typically, RML methods are more flexible with respect to the addition of new terms. However, they approximate the solution, instead of the more rigid primal-dual approach.
    \item We act on the hyper-parameters in a different way as the one done in RML methods. While RMLs need to survey the hyperparameter space to get the best combinations, we are only considering one hyperparameter in the cost functional for the time evolution, $\mu$ (and in a second instance, $\tau$). The flux and $\chi^2$ regularizers are Lagrange multipliers and thus,  their updating is a normal consequence of solving the optimization problem. This leads to a lower dimensional hyper-parameter space to be surveyed.
    \item \cite{mem} found the quasi-Newton step by using an approximation of the Hessian of the Lagrangian, composed by its tri-diagonal terms. This consideration is limiting in case of having a sparse array (as in the case of VLBI). Instead, in our implementation we work with the whole Hessian. As a consequence, we take into account the correct correlation among all pixels and thus we can better overcome the limitations imposed by having point spread functions (PSFs) with high and distant sidelobes, corresponding to sparse snapshot uv-coverages.

\end{itemize}

\section{Synthetic data I: Gaussian sources}\label{sec:synthetic_data}
\label{subsec:gaussian_sources}

In this section, we present the results obtained by applying the ngMEM algorithm to various synthetic data sets,{having different uv-coverages.} It is worth to recall that we understand by dynamic any kind of evolution/movement either in flux density, in relative position or in both. 

{Working on synthetic data from which we know the behavior of the sources in time, we can test the impact of several algorithmic choices and free parameters. Particularly we study the effect of $\mu$ in the movie reconstruction, and which role the kind of keyframe interpolation used plays, together with the visibility weighting scheme.}

The aim of this section is to solve the problem for different number of keyframes and present how the kind of keyframe interpolation used also plays an important role, together with the visibility weighting scheme.

\begin{figure}
    \centering
    \begin{subfigure}{0.5\textwidth}
        \includegraphics[width=\textwidth]{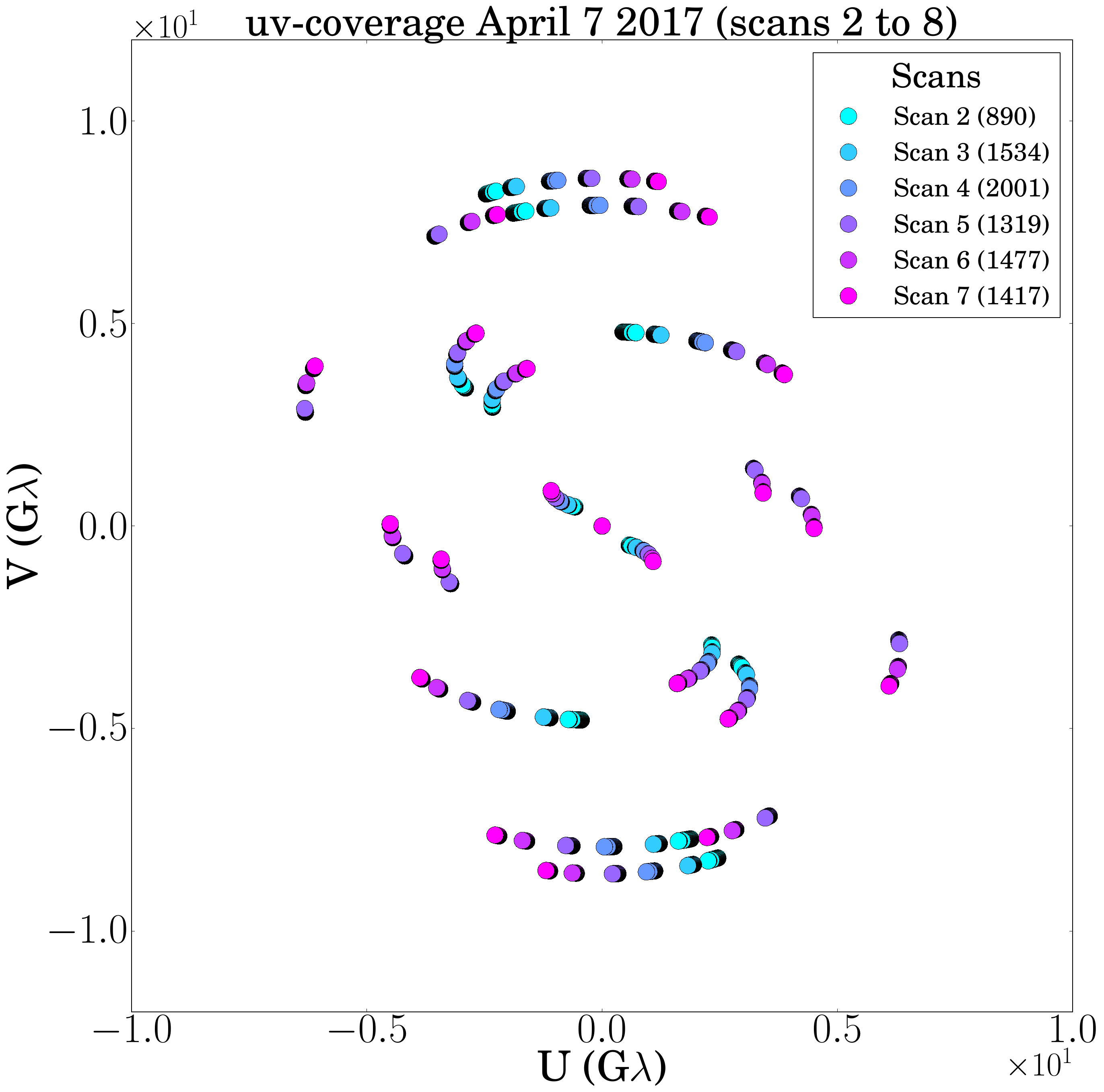}
        \label{fig:uvplot}
    \end{subfigure}
    \caption{uv-coverage of the EHT observations of \sgra, taken on April 7, 2017. The data are arranged in scans, shown in different colors. The number of visibilities in each scan are given in the figure legend (in parenthesis). }
    \label{fig:synth_all_cov}
\end{figure}

In a first instance,{ we have simulated four test sources} replicating the {array distribution and} observing times from April 11 of the 2017 EHT campaign,~(\citealt{EHT1} to \citealt{EHT4}), labeled as:

\begin{enumerate}
    \item \texttt{DOUBLE_1S:} A double source changing its brightness during the scan 2 (from 09:20:00.0 UT to 09:43:00.0 UT). During this scan, the uv-coverage was particularly bad (see Fig.~\ref{fig:synth_all_cov}). In this case, we want to test the algorithm acting on short and weak constrained problems.
    \item \texttt{FOUR_2S:} Four compact components, with two of them changing their brightness in time. The scans are 2 and 7 (the last one being from 12:48:00.0 UT to 13:09:00.0). Scan 7 had the best uv-coverage (Fig.~\ref{fig:synth_all_cov}). This fact, together with the big jump in time between scan 2 and scan 7, make this source a challenging test for the sensitivity of the algorithm to more complex flux-density evolution.
    \item \texttt{FOUR_FULL:} Same as point above, but with the scans 2 to 7.
    \item \texttt{DOUBLE_2S:} Same as first point, but for scans 2 and 7.
\end{enumerate}

The cases are selected based on their increasing level of difficulty. The initial case represents a straightforward scenario. In the second case, there is a significant temporal gap, leading to a substantial loss of uv-coverage. 

By examining this case, we demonstrate the beneficial impact of keyframes that contain more visibilities, as they aid in constraining the source brightness distribution during the reconstruction process for keyframes with fewer visibilities.

Finally, the last case involves an observation spanning approximately 10 hours, without any significantly large data gaps. We show how smoother keyframe divisions help for the reconstruction.\\
In all cases, the visibilities are contaminated by independent Gaussian noise in the real and the imaginary parts of the visibilities.

In this section, we have focused on two types of time interpolation methods for keyframes: nearest and linear interpolation. Nearest interpolation, also known as Nearest-Neighbor Interpolation, employs a piece-wise-constant function to interpolate pixel values between keyframes. On the other hand, linear interpolation provides a more accurate result by interpolating values in a linear manner. However, both methods may result in a lack of differentiability between adjacent keyframes, leading to a non-smooth evolution of the source.\\
{Future} developments will explore advanced interpolation techniques, such as those based on second-order polynomials or splines. Nevertheless, for the purpose of this section, the existing interpolation methods prove to be sufficient.\\
Moreover, for the sake of readability, the results presented in this section are obtained by utilizing the static ($\mu=0$) ngMEM reconstruction (solution) as the initial model for the entire experiment. In Appendix~\ref{sec:comparison_initial_points}, a comparison is provided by using either the dirty image or the snapshot image obtained using the static ngMEM as initial points for each of the four cases.

\subsection{Uniformly distributed keyframes}

{Choosing the appropriate (time) location of the keyframes, as discussed in Sect.~\ref{sec:ngmem_formalism}, is a crucial step.} The time distribution of the visibilities and how the times of the keyframes are set is a condition for the quality of the reconstructed movie. First, we show the results obtained by an equally spaced keyframe distribution (during the observing time). The total number of keyframes has been set to 3 and 4.

\begin{figure*}
    \hspace{-0.5cm}
    \includegraphics[scale=0.15]{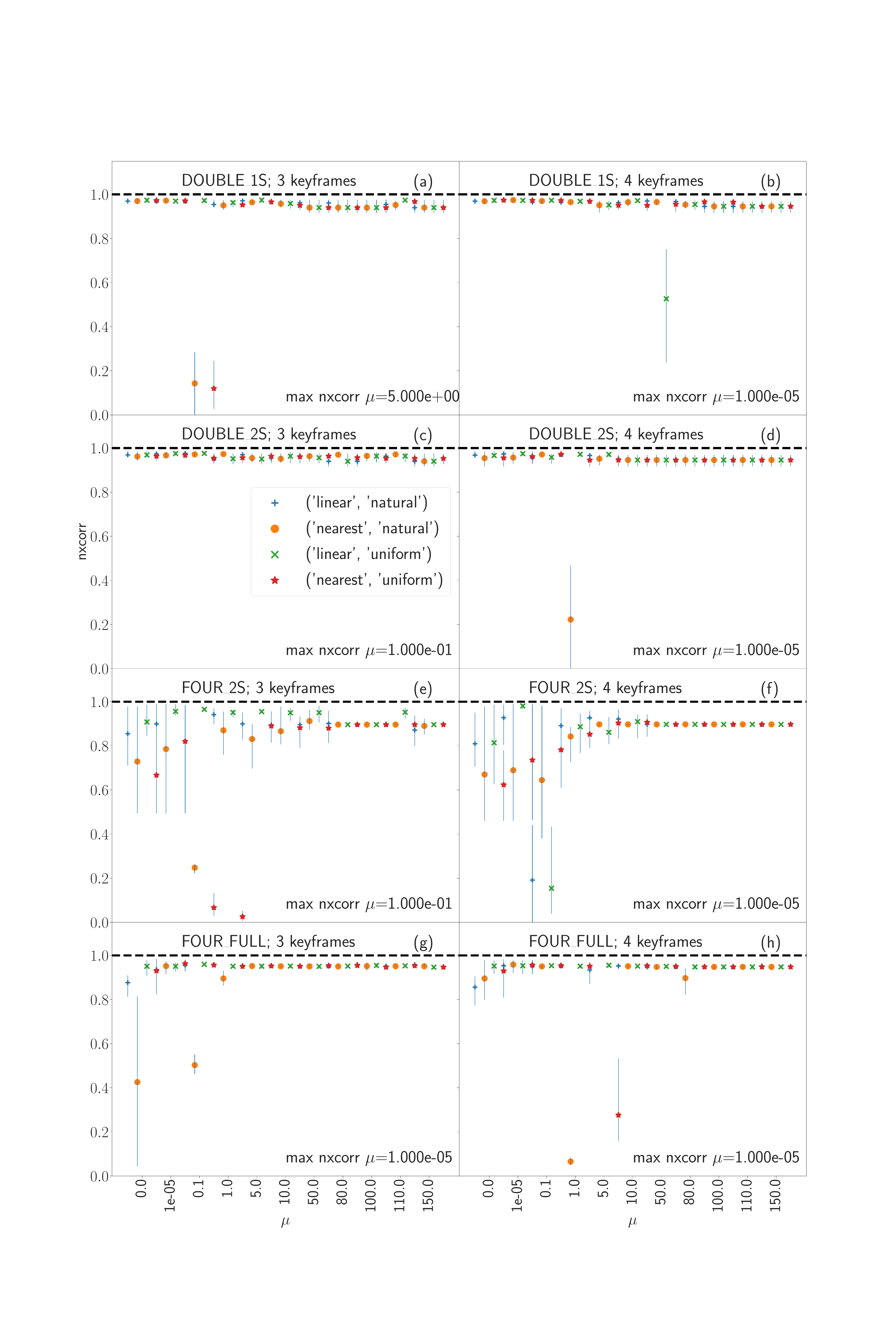}
    \vspace{-1cm}
    \caption{\nxcorr value for different $\mu$ for the four sources using three keyframes ((a) to (d)) and four keyframes ((e) to (h)). Different colors and markers represent different combinations of weighting and interpolation. In each plot, the $\mu$ producing the best movie reconstruction is indicated {(bottom right corner)}. Dash horizontal line indicates $\nxcorrm=1$. {For visual clarity and to avoid overlapping of the different markers, we have introduced an small offset in the $\mu$ axis.}}
    \label{fig:four_synth_dynm}
\end{figure*}

The figure of merit used to quantify the quality of the reconstruction has been the normalized cross-correlation, or \texttt{nxcorr},~\citep[see for instance][]{Farah22} between the true (time-dependent) source brightness distribution and the movie obtained using the ngMEM. This test measures the similarity of two (normalized) signals.\\
Figure~\ref{fig:four_synth_dynm} shows the effects of increasing $\mu$ with respect to the $\nxcorrm$.
The left four panels, namely (a) to (d), represent the observations of double sources obtained through one scan and two scans, utilizing three and four keyframes. On the other hand, the right four panels, denoted as (f) to (h), illustrate the corresponding results for the four Gaussian sources.

Different markers represent different combinations of interpolation type (linear/nearest) and weighting scheme (uniform/natural). The asymmetric error bars are the distance between the best keyframe (the one with highest \nxcorrm) and the worst with respect to the mean.

In this simple dataset, several {keytrends can be identified}: First, observe that, for all cases, $\mu=0$ has a high \nxcorr (with a higher variation for the case \texttt{FOUR_FULL} with near interpolation and natural weight). This is because these sources are morphologically simple. As we will see in the next sections, this is not the case for more complex source structures. Therefore, in these cases the algorithm can capture the variability just with the interpolation of the keyframes, if a proper keyframe division has been imposed. However, there is less variability between keyframes (in terms of error bars) when having more scans (see for instance Fig.~\ref{fig:four_synth_dynm}). This is because the scan with better uv-coverage helps to recover the evolution of the source. 

Second, in general and as expected, nearest interpolation is always worse than linear interpolation, although they are comparable within the error bars. This observation, along with others, has motivated us to develop a more accurate metric discussed later. 

\subsection{Keyframes chosen by observing times}

In a more general case, a uniform distribution of keyframes may not effectively capture the inherent variability of the source. To address this limitation, our algorithm provides the flexibility to strategically place keyframes that are specifically tailored to correspond with desired observing times. As a specific example, we have chosen to place a keyframe at the start and the end of each scan. This approach enables us to capture the evolution of the source within each scan. However, it should be noted that this method results in a larger number of keyframes, and therefore, an increased computational cost associated to the (much larger) Hessian of Eq.~\eqref{eq: hessian_time_dep} during the optimization process.

\begin{figure*}
    \centering
    \includegraphics[scale=0.3]{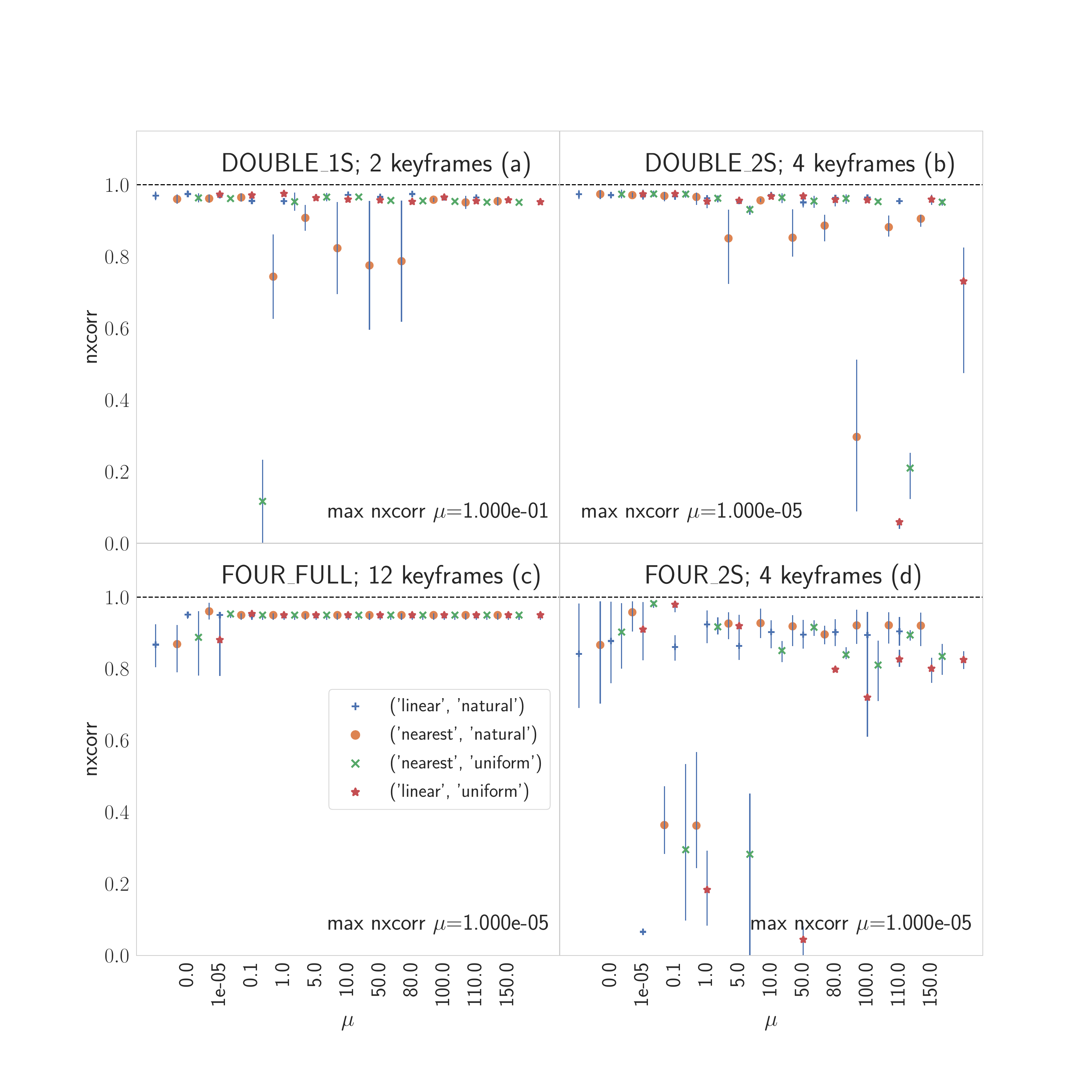}
    \vspace{-1cm}
    \caption{Same as Fig.~\ref{fig:four_synth_dynm}, but with keyframes set to the start and end of each scans (see Sect.~\ref{sec:synthetic_data}).}
    \label{fig:four_synth_dynm_myscans}
\end{figure*}

Figure~\ref{fig:four_synth_dynm_myscans} shows the \nxcorr versus $\mu$ for each simulation in the case when the keyframes are set to the start and end of each scan. 
It is worth remarking that we can see how the interpolation and weighting start to play a role, for example, for the case \texttt{DOUBLE_2S} or \texttt{FOUR_2S}. 

\subsection{Summary of the results}

In this section, we have applied the ngMEM algorithm to Gaussian-like synthetic sources using different scans of the EHT array in April 7. We have seen the effect of $\mu$ on movie reconstructions presenting different complexities.

The analysis highlighted the advantageous impact of keyframes with greater visibility counts in constraining source brightness distribution during reconstruction, especially for keyframes with fewer visibilities. We have verified the importance of the time interpolation (near and linear) between keyframes and how it helps to reconstruct a more better movie.

We have given two keyframe distribution strategies: uniformly across the scan and placing keyframes at the start and end of each scan. The keyframe distribution is crucial to capture the dynamics of the source.

Nevertheless, the kinematics of the source are simple enough to be easily captured. In the following section, we investigate different keyframe distributions in a more complex setting. In addition, we have noticed that the \nxcorr metric can be not accurate enough to evaluate the fidelity of a reconstruction and we will present a better figure of mertit.

\section{Synthetic data II: Ring-like structures. SgrA$^*$ plus a hotspot}\label{sec:synth_data_II}

\begin{figure}
    \centering
    \includegraphics[scale=0.2]{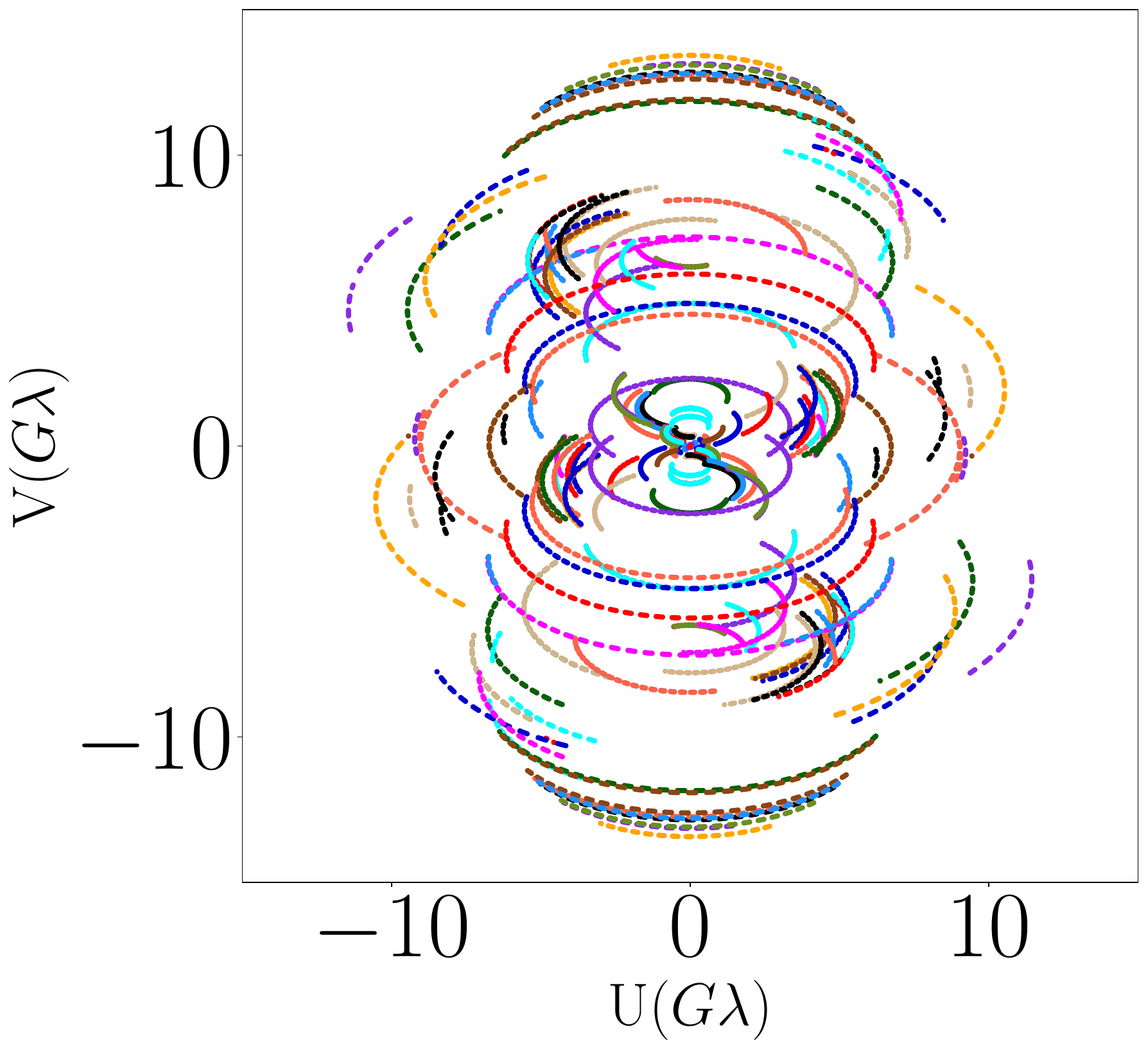}
    \caption{The ngEHT uv-coverage in a 24\,h observations of \sgra. Different colors represent different baselines. The 20 ngEHT antennas can be found in~\citef{Roelofs23}.}
    \label{fig:ng_eht_uv_cov_info}
\end{figure}

Using the EHT array extended with more antennas, called ngEHT array~\citep[e.g.][]{Roelofs23} we have simulated two different scenarios, at 345\,GHz, of a ring structure with an orbiting hotspot, each with distinct orbital periods. These simulations are intended to model a 24-hour observation of \sgra at one specific simulated epoch. The first simulation features a hotspot with a period of 24 hours {(we refer it as persistent hotspot)}, while the second simulation is based on the results of \cite{Wielgus22} {(or steady hotspot)}, which depicts a hotspot that is present for approximately 2 hours, completing two full orbits before dissipating. For the purpose of these simulations, we have disregarded any potential scattering effect~\citep[see][for more details on the noise modeling of \sgra]{EHT3,EHT4}, but we have added to the complex visibilities the thermal noise expected from the receiver and enhanced by atmospheric opacity based on~\cite{Roelofs23}.
Figure~\ref{fig:ng_eht_uv_cov_info} shows the coverage of the uv-plane. This array consist of 20 antennas,  with a shortest baseline of $\sim10^5\,\lambda$ and a longest baseline of $\sim$13.5\,G$\lambda$.

{The distinct time evolution of the source is crucial in determining the number of keyframes and their temporal distribution. This section highlights the major role these two factors play in successfully recovering the dynamics of the simulated hotspots.}

Moreover, as seen in the previous section, the $\nxcorrm$ metric is not enough to evaluate a movie. In this regard, we have developed the $\mfront$, a figure of merit that allows to evaluate the solution with respect to the model movie in every integration time. {This metric measures the distance between the obtained reconstruction and an ``ideal'' solution in every integration time. A smaller distance signifies a more accurate reconstruction (with respect to the ``ideal'').  We refer to Appendix~\ref{sec:mfront} for more details.}

Henceforth, we have fixed the weighting to be uniform and the time interpolation between two consecutive keyframes to be linear. Uniform weighting has been used to avoid possible over-weighting of the baselines associated to ALMA. The exceptional sensitivity of this telescope can result in the dominance of its visibilities within the image if a natural weighting scheme is applied. However, in the case of \sgra, this dominance can be used to derive a good estimate of $\fmod$ \citep[from intra-field dynamic amplitude calibration; see][]{Mus22, Wielgus22} to be used by the ngMEM. Hence, during the optimization process, the ``lightcurve constraint'' will force ${\displaystyle \sum_k}p^{l}_{k}$ to be as close as possible to $\fmod$, by updating the $\vec{\beta}$ Lagrange multiplier, just as the $\lambda$ multiplier will also be updated to enforce the $\chi^2$ condition. {To achieve this, we use},~Eq.~\eqref{eq:osculating_quadratic_2}.\\
In addition, the $\chi^2$ constraint {plays a }role defining the feasible set depending on the noise estimation. A too conservative estimation can result in an image with an incomplete deconvolution of the source structure, while a too low value of the imposed rms can introduce artifacts associated to noise. In this work, we estimate the visibility noise by computing the rms of the visibilities, divided in chunks of a given duration. We have tested different chunk periods for the two cases (constant and transient hotspot) to find the best noise estimate ($1.875\times 10^{-1}$ and $2.41\times 10^{-1}$ respectively).

In the next two sections, we show the obtained results for these two simulated cases.

\subsection{Persistent hotspot}

{For the case where the hotspot persists during the whole observation, the kinematics of the hotspot can be captured with a uniform distribution of keyframes. We have performed a gridsearch exploration, based on the \nxcorr metric, to find the number of keyframes and $\mu$ to best capture the movement.}  {In particular, we show the results} obtained using 10 and 15 keyframes. In the case of 10 keyframes, the optimal value found for the parameter $\mu$ is 700, whereas for 15 keyframes, $\mu$ is determined to be 200.

\begin{figure*}
    \centering
    \includegraphics[width=0.8\linewidth]{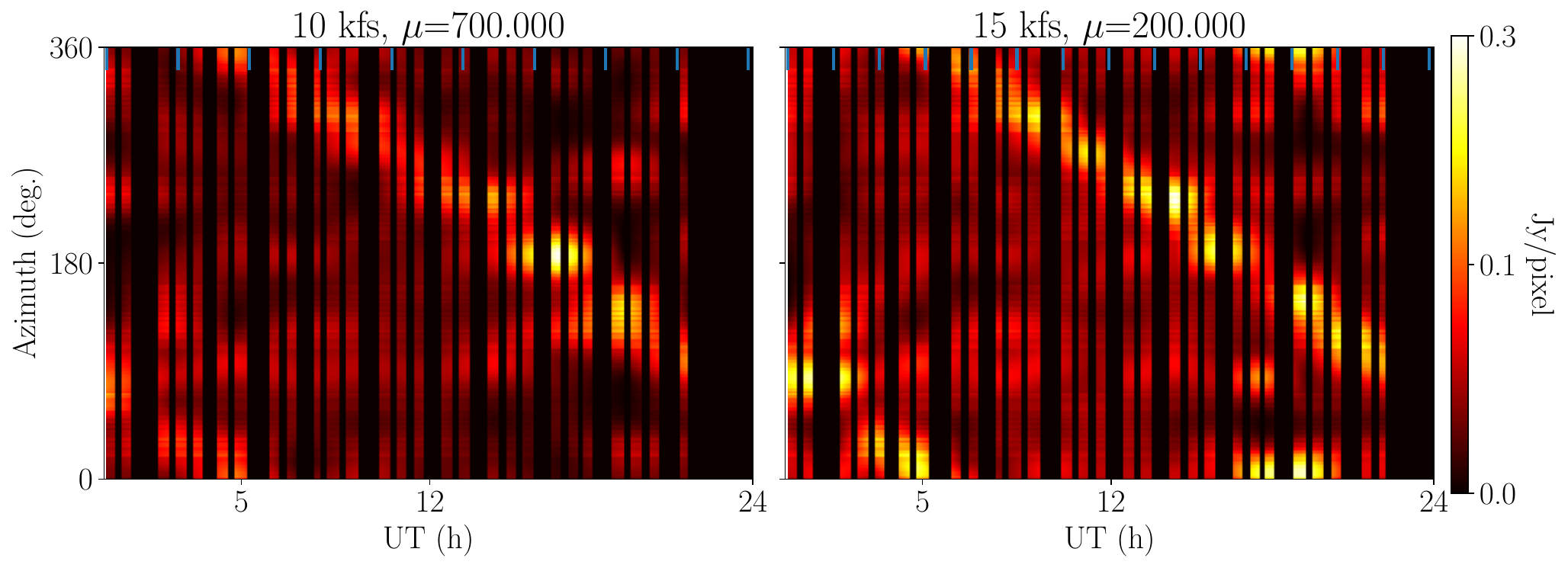}
    \caption{Azimuthal distribution of the ring brightness versus time for the persistent hotspot. Images between keyframes have been computed using a linear interpolation. Black stripes are placed where there are no visibilities and the vertical (blue) lines indicate the keyframe times. The movie is composed of 1024 frames and 10 keyframes. Color scale indicates brightness of the source. \textit{Left panel: $\mu=700$}. \textit{Right panel: $\mu=200$}.}
    \label{fig:phase_vs_frame_simple}
\end{figure*}

Figure~\ref{fig:phase_vs_frame_simple} illustrates the azimuth angle (vertical axis) and time (horizontal axis) corresponding to these optimal values. The figure showcases the reconstructed temporal variation of the azimuthal brightness distribution.\\
The hotspot brightness dominates over the noise and the ring. Since it is orbiting with constant angular velocity, the hotspot signal manifests as a diagonal feature in the figure. {We observe that as the number of keyframes increases, the definition of the hotspot's orbit becomes clearer.}

\subsection{Transient hotspot}

\begin{figure*}
    \centering
    \includegraphics[scale=0.29]{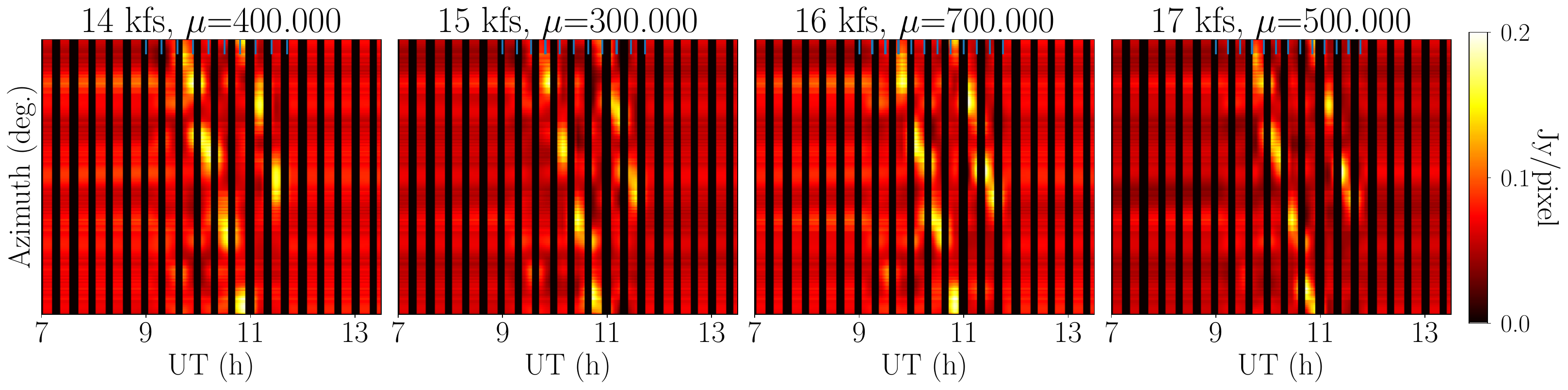}
    \caption{Azimuthal distribution of the ring brightness versus time for the transient hotspot model during the existence of the hotspot. Images between keyframes have been computed using a linear interpolation. Black stripes are placed where there are no visibilities and the vertical (blue) lines indicate the keyframe times.
    Color scale indicates brightness of the source.}
    \label{fig:phase_vs_frame_complex}
\end{figure*}

\begin{figure*}
    \centering
    \includegraphics[scale=0.29]{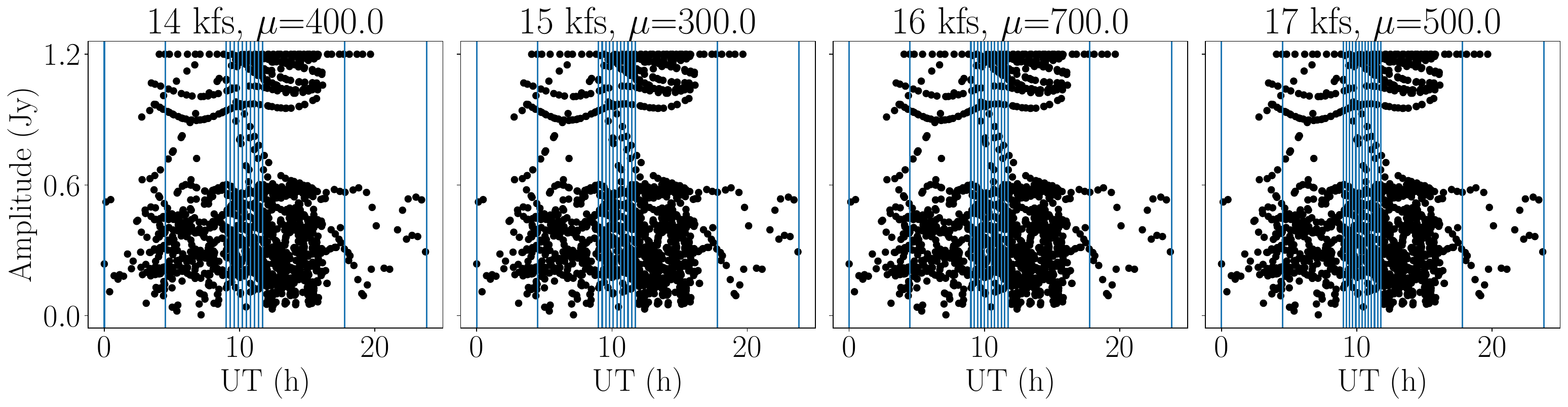}
    \\
    \includegraphics[scale=0.29]{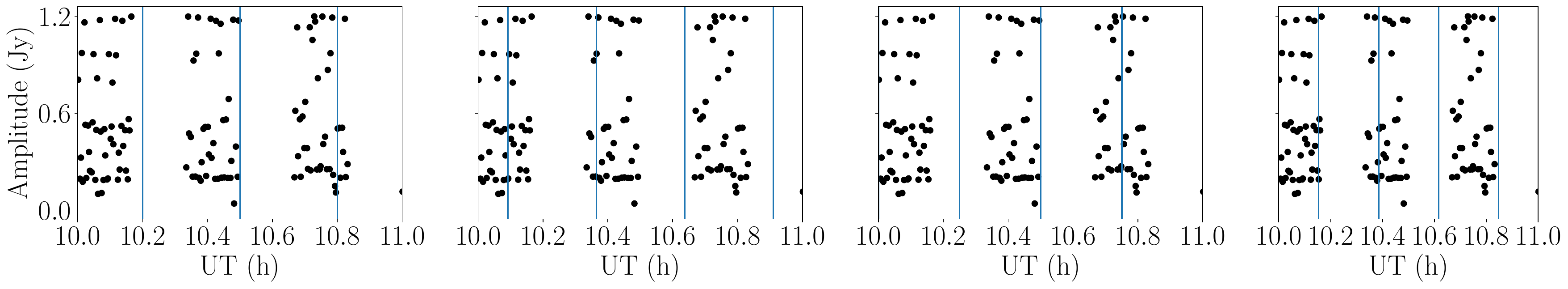}
    \caption{60s time-averaged visibility amplitudes (black dot points) corresponding to the transient hotspot simulation during the whole experiment. Keyframe times are shown as vertical blue lines. There are cases where the keyframes contain visibilities and cases where they do not. \textit{Bottom row}: Time interval zoom where there is a big gap on the visibilities. This corresponds to the discontinuity of the profile for the angles in the range $\sim$ 80 to 100 degrees.}
    \label{fig:visibilites_timeframes}
\end{figure*}

In contrast to the previous scenario, where scans were evenly distributed, we have manually selected a more strategic distribution for the scans in this case. Considering that the hotspot exists within a specific timespan of the observation, we have placed a higher density of keyframes during the presence of the hotspot and a lower density of keyframes over an extended timeframe when the hotspot is not present. This approach enables us to capture the movement of the hotspot more efficiently. {Moreover, in this case, we show how keyframe interpolation is useful when certain keyframe are placed in a time without associated visibility.}

{We have conducted a gridsearch using 14, 15, 16 and 17 keyframes. The results indicate that 15 keyframes produces the best movie based on the $\mfront$ figure of merit (see Appendix~\ref{sec:mfront})}. {We note, again, that increasing number of keyframes tends to recover a better orbit definition for the hotspot.}

In Fig.~\ref{fig:phase_vs_frame_complex}, we present profiles for four different numbers of keyframes: 14, 15, 16, and 17, displayed from right to left. Each profile is generated using the movie with a fixed value of $\tau$, representing the ``image memory'' as defined in Eq.~\eqref{eq:image_memory}. Specifically, we set $\tau = 0.08$ for 14 keyframes, $\tau = 0.07$ for 15 and 16 keyframes, and $\tau = 0.06$ for 17 keyframes (as indicated in Tab.~\ref{tab:distance_15_simple}), and $\mu$ set to the value that produces the best movie reconstruction, according to the $\mfront$.
Before $\sim$10:30 UT, a homogeneous brightness distribution, corresponding to the ring structure, can be seen. From $\sim$10:30 UT to $\sim$12:30 UT, which corresponds to $\sim$ 2\,h, we start to see the orbiting hotspot as a bright feature with a changing azimuthal angle with time.

{This highlights the importance of keyframe interpolation when certain keyframes lack associated visibilites.} Fig.~\ref{fig:visibilites_timeframes} shows the 60s averaged visibility amplitudes of the source as black dots along duration of the experiment (where its duration has been rescaled from 0 to 1) for the four cases (from left to right) of 14, 15, 16 and 17 keyframes. Vertical lines denote the position of the keyframes. If the vertical line intersects visibilities, the keyframe will have usable data at its central time. Otherwise, the model will time-interpolate and the $\tau$ parameter (i.e., the way information at different times is propagated to that keyframe) will have a major impact on the keyframe image reconstruction. The bottom set of panels is a zoom to the time-range from 10 UT to 11 UT, where some keyframes are not having any data point. This is translated into the ``discontinuity''  appearing in Fig.~\ref{fig:phase_vs_frame_complex} in around 80 to 100 degrees. Because this keyframe has visibilites relatively distant in time, the algorithm does not recover the source brightness with high accuracy around that time.

\subsection{Final remarks on synthetic data}

{Through Sec.~\ref{sec:synthetic_data} and Sec.~\ref{sec:synth_data_II}}, we have seen how the ngMEM algorithm is able to recover different types of kinematics for both simple sources, such as Gaussian profiles, and more complex structures like rings with an orbiting hotspot. Notably, in the case of ring structures with a hotspot, our algorithm effectively captures the evolution of the hotspot, regardless of whether it persists over time or undergoes rapid transitions.

To achieve optimal recovery of these kinematics, it is crucial to appropriately set the free parameters, which directly impacts the quality of the results. By selecting the optimal values for the free parameters, we can attain high scores in the figure of merit, indicating a successful reconstruction of the source's kinematics. In contrast with Sec.~\ref{subsec:gaussian_sources}, in all cases $\mu=0$ performs the worst (see Appendix~\ref{sec:mfront}). This fact proves that the time entropy regularizer helps to better capture the evolution of the source.

{Additionally, our algorithm demonstrates its capability to reconstruct movies even when isolated keyframes lack associated visibilities}. In such cases, the "image memory" parameter, $\tau$, {significantly influences} the reconstruction process. The value of $\tau$ influences how the algorithm uses information from neighboring keyframes to fill in gaps and ensure a coherent reconstruction, thereby mitigating the impact of missing visibilities.

\section{Comparison with another dynamical reconstruction method}
\label{sec:mj_comparison}

{As a conclusion of the results, we {qualitatively} compare ngMEM with  the RML method described in~\cite{Johnson17}.}
\cite{Johnson17} consider a particular instance of Prob.~\eqref{eq:dual_mem_timeindp_other_version} {which incorporates multiple dynamical regularizers. Particularly, they define $\mathcal{R}_{\Delta t}$, which aims to enforce that the brightness of the reconstructed image remains piecewise smooth between two adjacent snapshots, and $\mathcal{R}_{\Delta p}$, aiming to minimize the deviations of each snapshot from the mean image.}

In order to replicate the experiment conducted in their Fig. 4, we utilized the same dataset that was shared through private communication. This dataset was originally presented in~\cite{Shiokawa13} based on a 3D general relativity magneto-hydrodynamics (GRMHD) simulation.\\
In their reconstruction,~\cite{Johnson17} employed the dynamical regularizer $\mathcal{R}_{\Delta I}$ {(defined in their Eq. 8)}, suitable for cases in which frames present small variations with respect to the time-averaged image.

\begin{figure}
    \centering
    \includegraphics[width=0.8\linewidth]{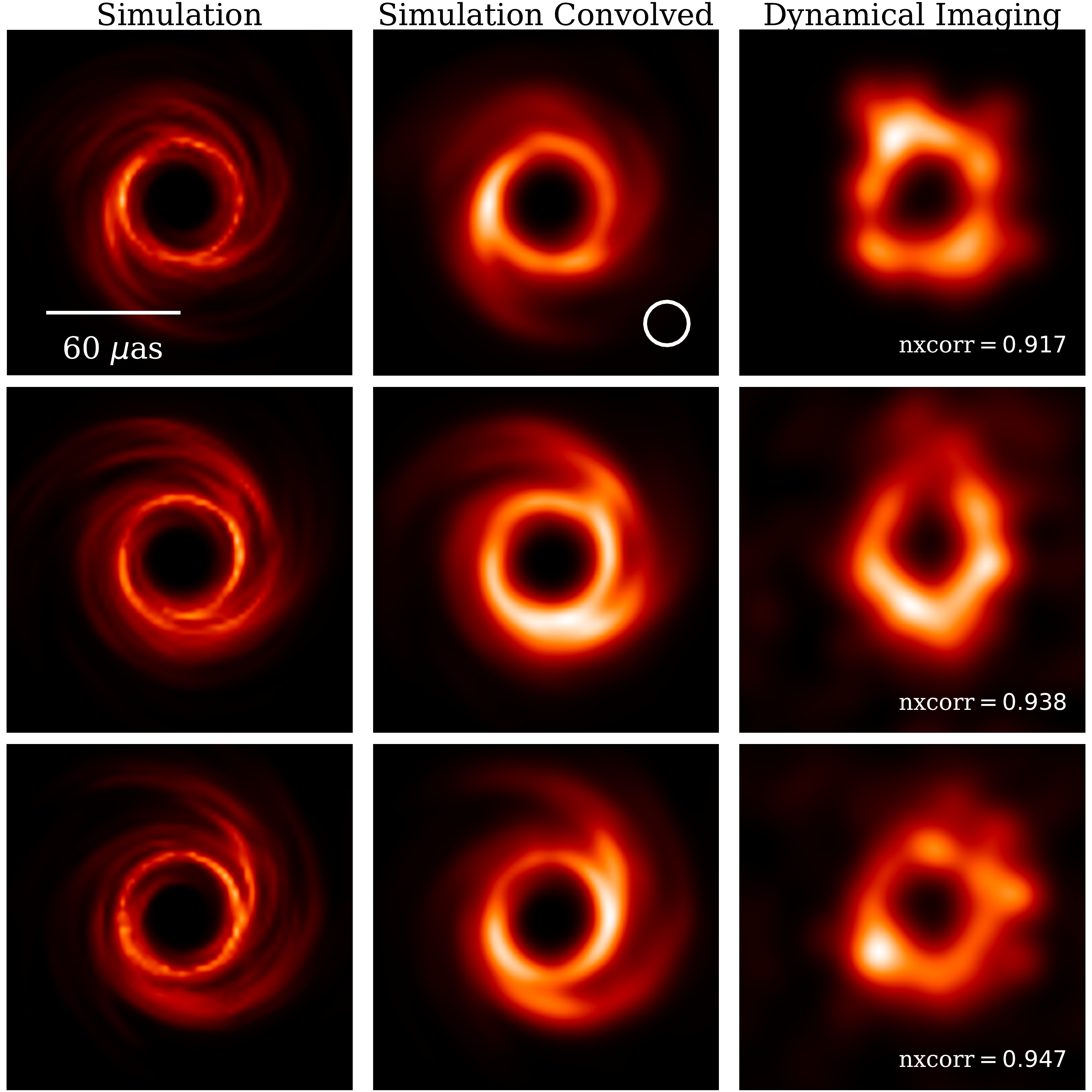}
    \caption{Example reconstruction of a face-on accretion disk presented in~\citef{Shiokawa13},~\citef{Johnson17}. The middle panels show the simulated images convolved with a $20\,\mu$as Gaussian and the right panels, dynamical reconstruction (using ngMEM) with $\mu=4$ convolved with the same beam. Each keyframe represents different times of the observation, but are consistent with the times appearing in Fig. 4~\citef{Johnson17}. The color scale is linear and is consistent among different times, but is scaled separately for each case, based on the maximum brightness over all frames. The \nxcorr value between the recovered and the convolved model can be found in the bottom corner of the reconstructed images. Data was provided by private communication.}
    \label{fig:mj_comparison}
\end{figure}

Figure~\ref{fig:mj_comparison} shows image reconstructions using ngMEM of the same model at the same times showed in the cited figure of~\cite{Johnson17} and $\mu=4$. The first column represents the model at three different moments (with three different instantaneous uv-coverage). The second column shows the model convolved with a $20\,\mu$as Gaussian. Third column is the ngMEM reconstruction. Because the early and late frames have few data constraints, the dynamical imaging reconstructions at those times do not recover the kinematics as good as for the middle keyframe.

Since~\cite{Johnson17}  do not report any quantitative comparison of their recovered keyframes with the true brightness distribution, it is not possible to compare quantitatively their results to ours. In any case, a visual inspection of Fig.~\ref{fig:mj_comparison} with Fig. 4 of~\cite{Johnson17} seems to indicates that ngMEM is performing better for this dataset.

The \nxcorr for every frame in our reconstruction can be seen in Fig.~\ref{fig:mj_comparison}.

\section{Summary and Conclusions}
\label{sec:summary}

Imaging in radio astronomy has long been a vibrant area of research, particularly in recent years with VLBI achieving unprecedented resolution through projects like the EHT. Furthermore, the future ngEHT promises significantly higher resolution capabilities, potentially enabling the temporal resolution of event-horizon scales of SMBHs with high intra-epoch time cadence.

Time variability has appeared to be a game-changing factor in the traditional conception of interferometric imaging, which proves to be critical for the fast-changing SMBH \sgra. This paper introduces a new non-convex, constrained optimization problem extending the traditional MEM formulated by~\cite{mem}, known as ngMEM. This extension involves integrating a new regularizer term into the cost functional to address time evolution. {It is the first MEM-type dynamic imaging algorithm proposed for VLBI.}

Our algorithm's main advantages are: 1) utilization of the full Hessian of the Lagrangian, leveraging information encoded in correlations among adjacent and distant pixels/keyframes, and 2) integration of a time entropy regularizer enforcing minimal time variability in the source image while remaining consistent with the data and {3.) a flexible framework for time interpolation and keyframe selection.}

{Along this paper, we have developed the primal-dual optimization formulation of the ngMEM. We have given a numerical method to solve it and we have justified its quadratic convergence. We have also listed the differences and advantages of this method with respect to some well-known alternatives, and we have proposed an iterative method to improve the starting point of the algorithm and, consequently, to get a better fidelity on the final movie.}\\
{We have tested this algorithm with different cases on synthetic data, even emulating a transient hotspot compatible with the results on \sgra reported by~\cite{Wielgus22}. To compare movie reconstructions and to quantify their fidelity, we have defined a novel figure of merit, the $\mfront$. Lastly, we have conducted a performance comparison between our algorithm and the algorithm proposed by~\cite{Johnson17}, by applying ngMEM to the same dataset they presented in Figure 4 of their study.}

{This algorithm can be easily adapted to perform dynamic studies, not only of \sgra and other active galactic nuclei, but also X-ray binaries, inter-epoch of long term observations and many other cases concerning kinematics. Currently, we are in the process of extending this algorithm to include antenna gains and polarimetry, with the aim of being able to recover robust polarimetric movies of fast-changing radio sources.}

\section*{Acknowledgements}
This work has been supported by the MICINN Research Project PID2019-108995GB-C22. IMV and AM also thank the Generalitat Valenciana for funding, in the frame of the GenT Project CIDEGENT/2018/021.\\
A.M. acknowledge Michael Johnson for sharing the scripts and images to recover their solution presented in their paper and Teresa Toscano (iaa-csic) for the proofreading of the manuscript.

\section*{Data availability}
The data, code, and other visual materials underlying this article are available in \textit{ngMEM_constrained}, at \url{https://github.com/AlejandroMus/ngMEM_constrained.git}.

\input{Appendix_Mfront.tex}
\input{Appendix_movies.tex}
\input{Appendix_initial_points}
\input{Appendix_static}

\end{document}

%% file: Appendix_Mfront.tex
\appendix
\section{A stronger criterion to compare movies}
\label{sec:mfront}

The figure of merit based on asymmetric error bars gives a good overall idea of the behavior of the reconstruction with different parameter combination, such as $\mu$, the visibility weighting scheme or the time interpolation. However, it is not a very accurate metric for selecting the better movie, since among other problems, in general, big error bars can be obtained, 
and therefore a big subset of solutions could be congruent within the error bars. For this reason, we have developed a new metric explained in this appendix.

\subsection{Formalism}

\begin{remark}
Throughout this section we suppose we have a groundtruth that we use to compare the \nxcorr of the different movies.
\end{remark}

In this appendix, we denote by $\kappa$ a movie reconstruction\footnote{In Greek, the word ``frame'' is \textgreek{καρέ}. Since a movie is a set of keyframes, we have decided to denote a movie by the first letter $\kappa$ of the word \textgreek{καρέ} to avoid confusion with the index $m$ appearing in the main text.}, i.e., \textit{a solution of the ngMEM}. Any $\kappa$ is composed by a ordered set of keyframes. So, given a movie reconstructed with the ngMEM, $\kappa$, of $n$ keyframes, we can write $\kappa:=\left\{f^{\kappa}_{1},\ldots,f^{\kappa}_{n}\right\}$. We denote by $\nxcorrm\left(f^{\kappa}_{k}\right),\ 1\leq k\leq n$ the \nxcorr of the $k$-th keyframe.\\
In case of having a set of $p$ movies composed by $n$ keyframes each, we denote by $f^{\kappa_l}_{j}$ the keyframe $j$ corresponding to the movie $\kappa_l, 1\leq l\leq p$.  

\begin{definition}[Set of movies]
{Given a fixed set of visibilities corresponding to a specific observation, and a fixed number of keyframes, we denote by $\mathcal{K}$ the set of all local minima (by setting different parameters, as $\mu$) of the ngMEM. Every element $\kappa\in\mathcal{K}$ is a movie.}
\end{definition}

\begin{definition}[Dominant keyframe]
A keyframe $f^{\kappa_i}_{l}$ of a movie $\kappa_i\in \mathcal{K}$ is dominant over a keyframe $f^{\kappa_j}_{l}$ of a movie $\kappa_j\in \mathcal{K}$  if, and only if, $\nxcorrm\left(f^{\kappa_i}_{l}\right)>\nxcorrm\left(f^{\kappa_j}_{l}\right)$.
\end{definition}

\begin{definition}[Super-dominant keyframe]
Given a movie $\hat{\kappa}\in \mathcal{K}$ we say its keyframe $f^{\hat{\kappa}}_i$ is super-dominant if $\nxcorrm\left(f^{\hat{\kappa}}_{i}\right)>\nxcorrm\left(f^{\kappa_l}_{i}\right),\ \forall \kappa_l\in\mathcal{K}$.
\end{definition}

\noindent Further elaborating on this concept, suppose that the set $\mathcal{K}$ consists of $l$ movies. Observe that, for all movies in $\mathcal{K}$, we can fix one keyframe. Then, since the set of keyframes $\left\{f^{\kappa_1}_1,\ldots,f^{\kappa_1}_n\right\}$ is a totally ordered set with respect to the \nxcorr metric, we can pick the best keyframe (in terms of \nxcorrm) among all movies. This chosen keyframe, is a super-dominant frame.

\begin{definition}[Dominant solution]
We say that a movie $\hat{\kappa}\in\mathcal{K}$ is a dominant solution, or a dominant movie, over a movie $\kappa_l\in\mathcal{K}$ if all its keyframes are dominant keyframes over $\kappa_l$. In other words, if $\hat{\kappa}:=\left\{f^{\hat{\kappa}}_{1},\ldots,f^{\hat{\kappa}}_{n}\right\}$ and $ \kappa_l:=\left\{f^{\kappa_l}_{1},\ldots,f^{\kappa_l}_{n}\right\}$, all their keyframes satisfies that $\nxcorrm\left(f^{\hat{\kappa}}_{i}\right)>\nxcorrm\left(f^{\kappa_l}_{i}\right),\ \forall i=1,\ldots,n.$
\end{definition}

\begin{definition}[$\mfront$]
The set $\left\{\hat{f}_1,\ldots,\hat{f}_n\right\}$ where each $\hat{f}_i,\ i=1,\ldots,n$ is a super-dominant keyframe is called the $\mfront$.
\end{definition}

\noindent This set determines an upper bound for the set keyframes. None of the movies in $\mathcal{K}$ have a frame better than the corresponding one of the $\mfront$. Of course, the movie generated by joining such a keyframes in general is not a valid movie, meaning \textit{is not, necessarily, a solution of the ngMEM}, but it would be the best, the desired, solution to get. Only if the ngMEM has a global solution and it is in $\mathcal{K}$ or $\mathcal{K}$ is a singleton, then it would happen that the $\mfront$ is a feasible movie.

\begin{definition}[Ideal movie]
The movie (assuming the abuse of language, since it is not a movie understood as a solution of the ngMEM) generated by the keyframes of the $\mfront$ is called the ideal movie and is denoted by $\tilde{\kappa}$.
\end{definition}

\begin{remark}
As stated above, in general $\tilde{\kappa}\notin\mathcal{K}$.
\end{remark}

\noindent The name of ideal is well-motivated. In the best case scenario, we would get the $\mfront$ movie. However, because of data and algorithm limitations, we never recover the groundtruth model. We will never get a movie where all keyframes have \nxcorr 1. But, in order to define the metric, we need this groundtruth. 

\begin{definition}[Utopian movie]
The groundtruth movie is called the Utopian movie and it's denoted by $\check{m}$.
\end{definition}

\noindent Now we are in the setting of establishing a criterion to compare two movie reconstructions, by defining their distance to the front.

\begin{definition}[Distance between two movies]
Given two movies from $\mathcal{K}$, $\kappa_1 := \left\{f^{\kappa_1}_1,\ldots,f^{\kappa_1}_n\right\}$ and $\kappa_2 := \left\{f^{\kappa_2}_1,\ldots,f^{\kappa_2}_n\right\}$, the distance between them is given by
\begin{equation}\label{eq:eq_kfront}
    d\left(\kappa_1,\kappa_2\right) := \displaystyle{\sum_{i}^{n}}\lvert \nxcorrm\left(f^{\kappa_1}_{i}\right) - \nxcorrm\left(f^{\kappa_2}_{i}\right)\rvert.
\end{equation}
\end{definition}

\noindent Observe that $d\left(\kappa_1,\kappa_2\right)$ defines a mathematical distance: it is positive definite, symmetric and satisfies the triangular inequality.\\
This distance tells us how distant are $\kappa_1$ and $\kappa_2$. As bigger the distance is, the less similar the movies are. Then, in order to get a criterion to chose the best solution among the possible reconstruction, we just need to compute the distance to the $\mfront$.

\begin{definition}[Distance to the front]
The best possible movie in $\mathcal{K}$ is the one whose distance to the front (or to the Utopian movie) is minimal. Such a movie is denoted by $\hat{\kappa}$ .
\end{definition}

\begin{remark}
For all movie $\kappa\in\mathcal{K}$, it always holds
\begin{equation}
    d\left(\tilde{\kappa},\check{\kappa}\right) < d\left(\check{\kappa},\hat{\kappa}\right) \leq  d\left(\hat{\kappa},\kappa\right).
\end{equation}
Since the model remains independent of the number of keyframes (each keyframe have \nxcorr one), movies with different keyframe distribution can be compared by normalizing the distance to the Utopian through division of each term in summation Eq.~\eqref{eq:eq_kfront} by the respective keyframe length.
The distance to the Utopian movie serves for comparing movies ``interkeyframe'', while the distance to the $\mfront$ provides a relative comparison to other solutions obtained in the search of the parameter space (``intrakeyframe''). Utilizing the $\mfront$ as a guide can aid in determining if further exploration of the parameter space is necessary.
\end{remark}

With this framework, we can find the best reconstruction (the one with less distance to the $\mfront$) for the more complex source: ring plus a hotspot and for the ring with a hotspot simulating the one of \sgra.

\subsection{$\mfront$ applied to the Synthetic data II}

\begin{figure*}
    \centering
    \includegraphics[scale=0.4]{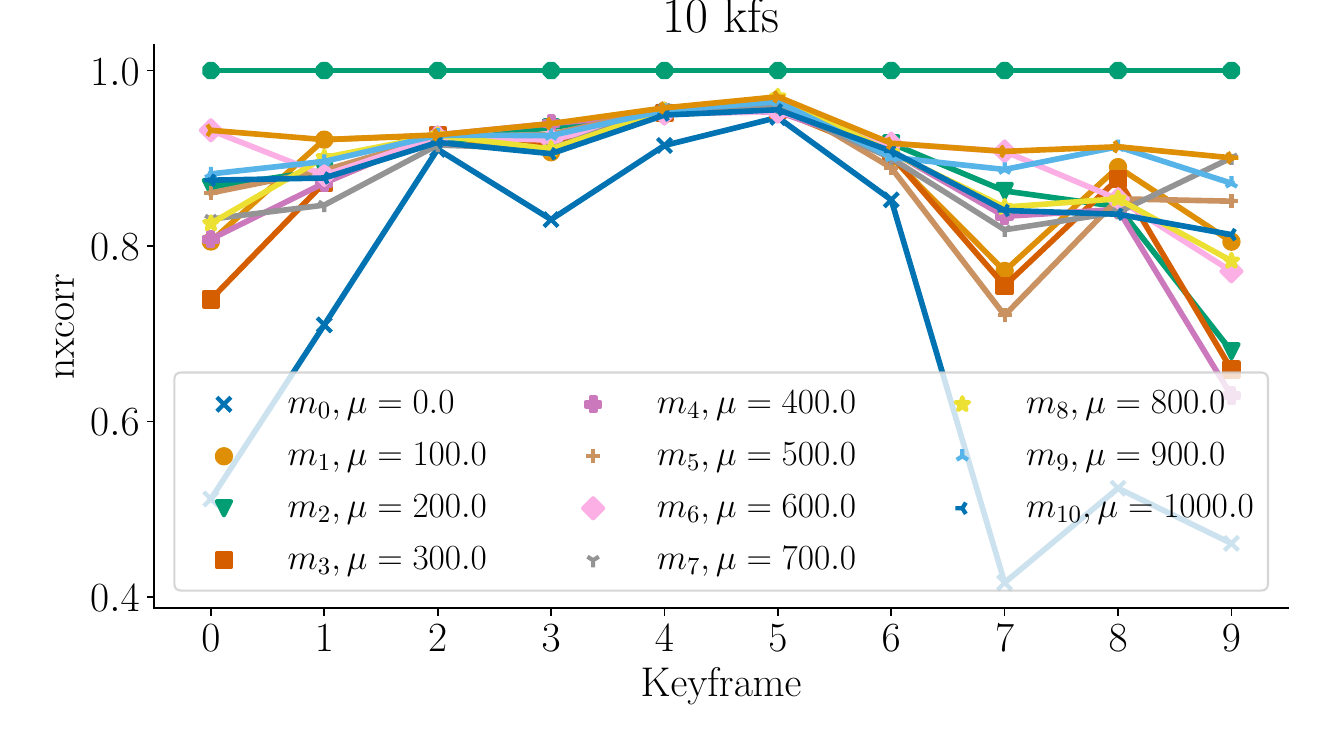}
    \includegraphics[scale=0.4]{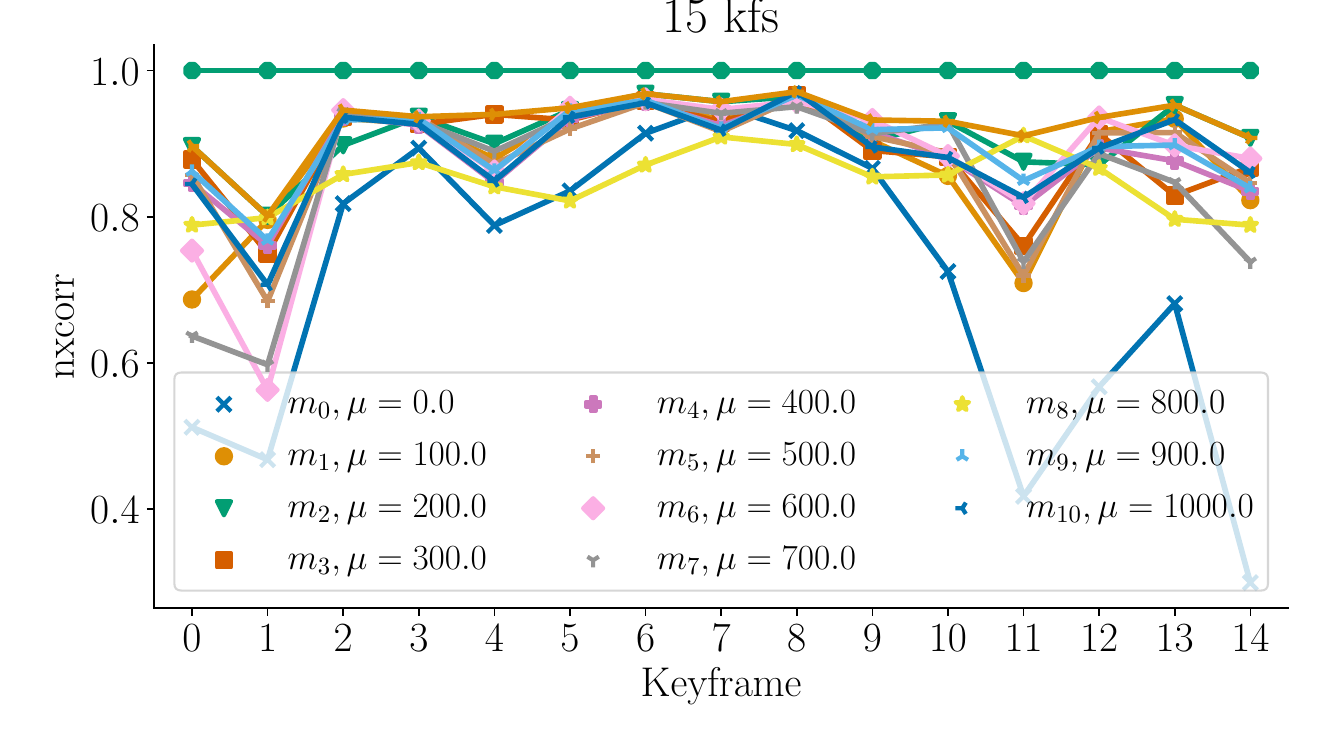}
    \caption{Values of the \nxcorr for every keyframe for different $\mu$. Left panel is for the case of 10 keyframes and right panel for 15 keyframes}
    \label{fig:mfront_10_superslow}
\end{figure*}
\begin{figure*}
    \centering
    \includegraphics[scale=0.4]{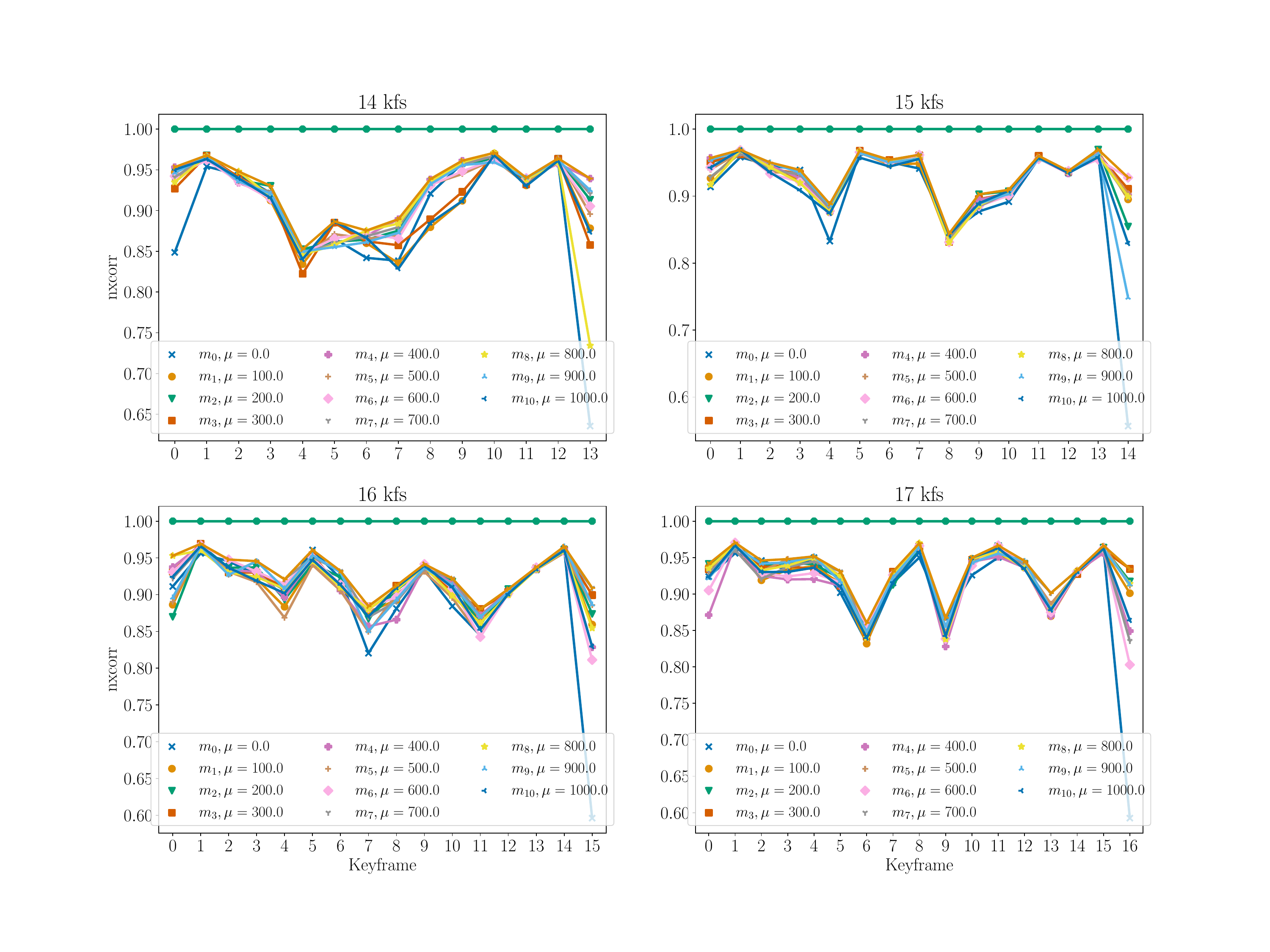}
    \caption{Comparison between 10 keyframes movies for different values of $\mu$ for the ring and hotspot appearing for 2\,h. \textit{Top left panel}: 14 keyframes. \textit{Top right panel}: 15 keyframes. \textit{Bottom left panel}: 16 keyframes. \textit{Bottom right panel}: 17 keyframes}
    \label{fig:mfront_complex}
\end{figure*}

In Fig.~\ref{fig:mfront_10_superslow}, we show the space $\mathcal{K}$
for the case of ring with a constant period hotspot. The x-axis of the figure are keyframes (and the keyframes are labeled with their associated number) and the y-axis the \nxcorrm. The movie with constant \nxcorr equals to 1 is the Utopian movie (the perfect/groundtruth movie). The ideal movie is also plotted and it is the upper bound frame-wise of the all movies. In Tab.\ref{tab:distance_10_superslow} we can find the distance of every reconstruction to the ideal and Utopian movie for 10 and 15 keyframes. Among all solutions, the closer one (the best) is for the case $\mu=700$ and $\mu=200$ respectively.

In Fig.~\ref{fig:mfront_complex}, we show the set of movies $\mathcal{K}$ for the case of \sgra and the spurious hotspot for 14 to 17 keyframes (top left, top right, bottom center resp.). Tables~\ref{tab:all_distances} show the distance to front for all this cases. The profiles presented in Fig.~\ref{fig:phase_vs_frame_complex} are done with movies with less distance.

With these examples, we can see the potential of the $\mfront$: visually we can get an idea of the better movies and easily find them without any uncertainty or ambiguity.\\
Again, we can observe how $\mu=0$ is the movie performing the worst. Time entropy always helps to recover the variability of the source.

\begin{table*}
\caption{Distance to the $\mfront$ for each movie reconstruction}
\centering
\begin{tabular}{lrrr}
\toprule
\small movie ($\kappa$) & \small $\mu$ & \small $d\left(\tilde{\kappa},\kappa\right)$ & \small $d\left(\check{\kappa},\kappa\right)$ \\
\midrule
    \small movie\_0 & \small 0.0 & \small 2.92275 & \small 2.20873 \\
    \small movie\_1 & \small 100.0 & \small 1.16771 & \small 0.45369 \\
    \small movie\_2 & \small 200.0 & \small 1.18200 & \small 0.46798 \\
    \small movie\_3 & \small 300.0 & \small 1.44435 & \small 0.73033 \\
    \small movie\_4 & \small 400.0 & \small 1.33665 & \small 0.62263 \\
    \small movie\_5 & \small 500.0 & \small 1.16342 & \small 0.44940 \\
    \small movie\_6 & \small 600.0 & \small 0.98863 & \small 0.27461 \\
    \small movie\_7 & \small 700.0 & \small 1.12293 & \small 0.40891 \\
    \small movie\_8 & \small 800.0 & \small 1.12393 & \small 0.40991 \\
    \small movie\_9 & \small 900.0 & \small 0.87672 & \small 0.16270 \\
    \small movie\_10 & \small 1000.0 & \small 1.12119 & \small 0.40717 \\
\bottomrule
\label{tab:distance_10_superslow}
\end{tabular}
\quad
\begin{tabular}{lrrr}
\toprule
\small movie ($\kappa$) & \small $\mu$ & \small $d\left(\tilde{\kappa},\kappa\right)$ & \small $d\left(\check{\kappa},\kappa\right)$ \\
\midrule
    \small movie\_0 & \small 0.0 & \small 4.34699 & \small 3.28353 \\
    \small movie\_1 & \small 100.0 & \small 1.84073 & \small 0.77727 \\
    \small movie\_2 & \small 200.0 & \small 1.29014 & \small 0.22668 \\
    \small movie\_3 & \small 300.0 & \small 1.63894 & \small 0.57548 \\
    \small movie\_4 & \small 400.0 & \small 1.68981 & \small 0.62635 \\
    \small movie\_5 & \small 500.0 & \small 1.76373 & \small 0.70027 \\
    \small movie\_6 & \small 600.0 & \small 1.78340 & \small 0.71994 \\
    \small movie\_7 & \small 700.0 & \small 2.16697 & \small 1.10351 \\
    \small movie\_8 & \small 800.0 & \small 2.26580 & \small 1.20234 \\
    \small movie\_9 & \small 900.0 & \small 1.52551 & \small 0.46205 \\
    \small movie\_10 & \small 1000.0 & \small 1.66490 & \small 0.60144 \\
\bottomrule
\label{tab:distance_15_superslow}
\end{tabular}
\\
\textbf{Notes:} Distance to the $\mfront$ of ring with a persistent hotspot. \textit{Left table}: 10 keyframes. \textit{Right table}: 15 keyframes.
\end{table*}

\begin{table*}
\caption{Distance to the $\mfront$ for each movie reconstruction}
\centering
\begin{tabular}{lrrr}
\toprule
movie ($\kappa$) & $\mu$ & $d\left(\tilde{\kappa},\kappa\right)$ &  $d\left(\check{\kappa},\kappa\right)$   \\
\midrule
  movie\_0 &     0.0 &                  1.56242 &                0.58004 \\
  movie\_1 &   100.0 &                  1.28623 &                0.30385 \\
  movie\_2 &   200.0 &                  1.10084 &                0.11846 \\
  movie\_3 &   300.0 &                  1.28322 &                0.30084 \\
  movie\_4 &   400.0 &                  1.06225 &                0.07987 \\
  movie\_5 &   500.0 &                  1.15602 &                0.17364 \\
  movie\_6 &   600.0 &                  1.14874 &                0.16636 \\
  movie\_7 &   700.0 &                  1.09573 &                0.11335 \\
  movie\_8 &   800.0 &                  1.27293 &                0.29055 \\
  movie\_9 &   900.0 &                  1.11730 &                0.13492 \\
 movie\_10 &  1000.0 &                  1.27924 &                0.29686 \\
\bottomrule
\label{tab:distance_14}
\end{tabular}
\quad
\begin{tabular}{lrrr}
\toprule
movie ($\kappa$) & $\mu$ & $d\left(\tilde{\kappa},\kappa\right)$ &  $d\left(\check{\kappa},\kappa\right)$  \\
\midrule
  movie\_0 &     0.0 &                  1.53114 &                0.56825 \\
  movie\_1 &   100.0 &                  1.12860 &                0.16571 \\
  movie\_2 &   200.0 &                  1.09935 &                0.13646 \\
  movie\_3 &   300.0 &                  1.04542 &                0.08253 \\
  movie\_4 &   400.0 &                  1.06920 &                0.10631 \\
  movie\_5 &   500.0 &                  1.06972 &                0.10683 \\
  movie\_6 &   600.0 &                  1.08818 &                0.12529 \\
  movie\_7 &   700.0 &                  1.09681 &                0.13392 \\
  movie\_8 &   800.0 &                  1.12506 &                0.16217 \\
  movie\_9 &   900.0 &                  1.19636 &                0.23347 \\
 movie\_10 &  1000.0 &                  1.19526 &                0.23237 \\
\bottomrule
\label{tab:distance_15}
\end{tabular}
\\
\begin{tabular}{lrrr}
\toprule
movie ($\kappa$) & $\mu$ & $d\left(\tilde{\kappa},\kappa\right)$ &  $d\left(\check{\kappa},\kappa\right)$   \\
\midrule
  movie\_0 &     0.0 &                  1.71033 &                0.59962 \\
  movie\_1 &   100.0 &                  1.42663 &                0.31592 \\
  movie\_2 &   200.0 &                  1.36647 &                0.25576 \\
  movie\_3 &   300.0 &                  1.24689 &                0.13618 \\
  movie\_4 &   400.0 &                  1.40190 &                0.29119 \\
  movie\_5 &   500.0 &                  1.47790 &                0.36719 \\
  movie\_6 &   600.0 &                  1.36490 &                0.25419 \\
  movie\_7 &   700.0 &                  1.24205 &                0.13134 \\
  movie\_8 &   800.0 &                  1.33694 &                0.22623 \\
  movie\_9 &   900.0 &                  1.32049 &                0.20978 \\
 movie\_10 &  1000.0 &                  1.38368 &                0.27297 \\
\bottomrule
\label{tab:distance_16}
\end{tabular}
\quad
\begin{tabular}{lrrr}
\toprule
movie ($\kappa$) & $\mu$ & $d\left(\tilde{\kappa},\kappa\right)$ &  $d\left(\check{\kappa},\kappa\right)$   \\
\midrule
  movie\_0 &     0.0 &                  1.64760 &                0.56044 \\
  movie\_1 &   100.0 &                  1.32389 &                0.23673 \\
  movie\_2 &   200.0 &                  1.23608 &                0.14892 \\
  movie\_3 &   300.0 &                  1.24140 &                0.15424 \\
  movie\_4 &   400.0 &                  1.47708 &                0.38992 \\
  movie\_5 &   500.0 &                  1.20981 &                0.12265 \\
  movie\_6 &   600.0 &                  1.42261 &                0.33545 \\
  movie\_7 &   700.0 &                  1.29893 &                0.21177 \\
  movie\_8 &   800.0 &                  1.24809 &                0.16093 \\
  movie\_9 &   900.0 &                  1.24040 &                0.15324 \\
 movie\_10 &  1000.0 &                  1.36439 &                0.27723 \\
\bottomrule
\label{tab:distance_17}
\end{tabular}
\\
\label{tab:all_distances}
\textbf{Notes:} Distance to the $\mfront$ of ring with a hotspot appearing only for 2\,h. \textit{Top left table}: 14 keyframes. \textit{Top right table}: 15 keyframes. \textit{Bottom left table}: 16 keyframes. \textit{Bottom right table}: 17 keyframes. The lesser the distance, the better the reconstruction.
\end{table*}

%% file: Appendix_movies.tex
\section{Movies}

In this Appendix we show the best movie reconstruction (in terms of the $\mfront$) together with the optimal values or $\mu$ with the associated $\tau$. The matrix of plots should be read from top to bottom and from right to left.\\
For the case of the persistent hotspot, the dispersion of the recovered visibility amplitudes is 1.5\%.

\begin{figure*}
\centering
\begin{subfigure}{0.5\textwidth}
  \centering
  \includegraphics[width=0.8\linewidth]{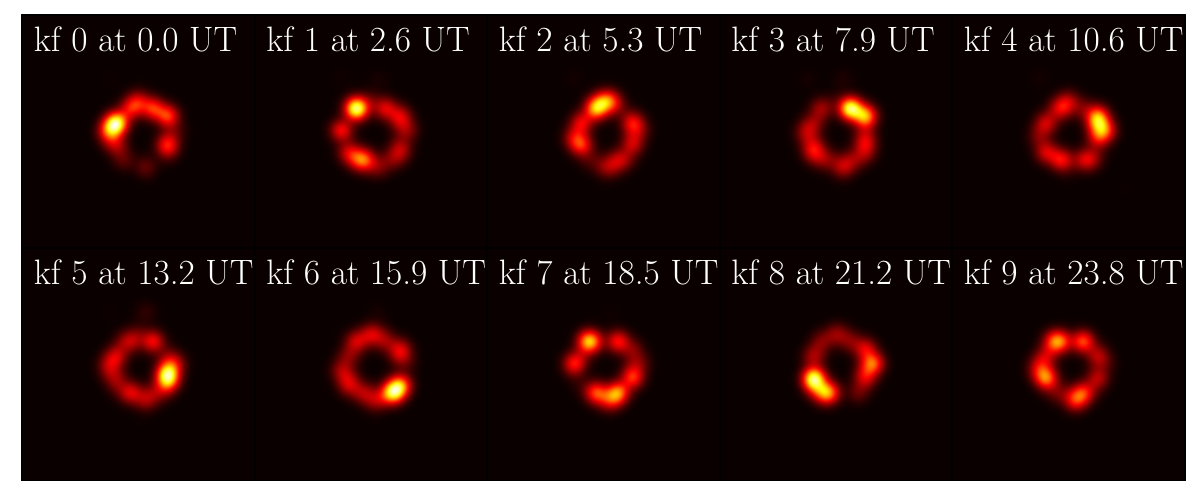}
  \vspace{-.15cm}
  \caption{\sgra plus a persistent hotspot, 10 keyframes}
\end{subfigure}%
\begin{subfigure}{0.5\textwidth}
  \centering
  \includegraphics[width=0.4\linewidth]{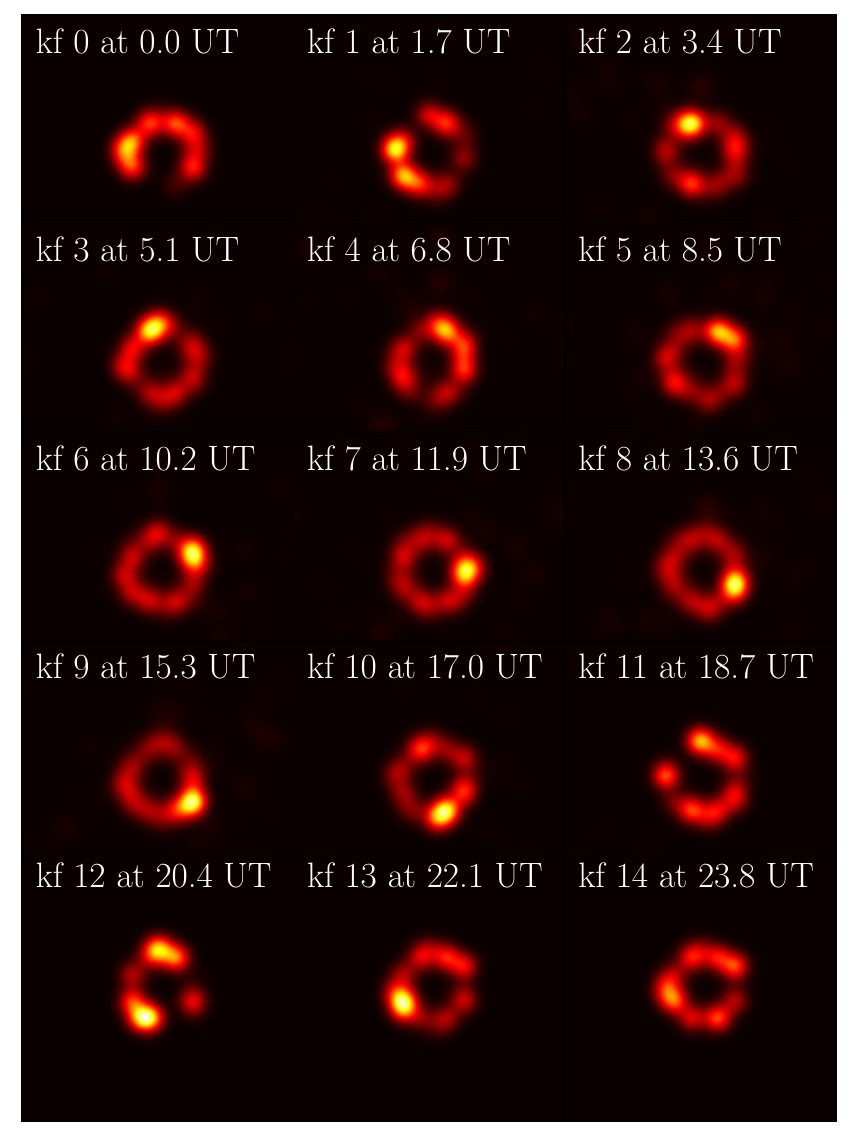}
  \vspace{-.15cm}
  \caption{\sgra plus a persistent hotspot, 15 keyframes}
  \label{fig: crescent_eht}
\end{subfigure}

\begin{subfigure}{0.5\textwidth}
  \centering
  \includegraphics[width=0.8\linewidth]{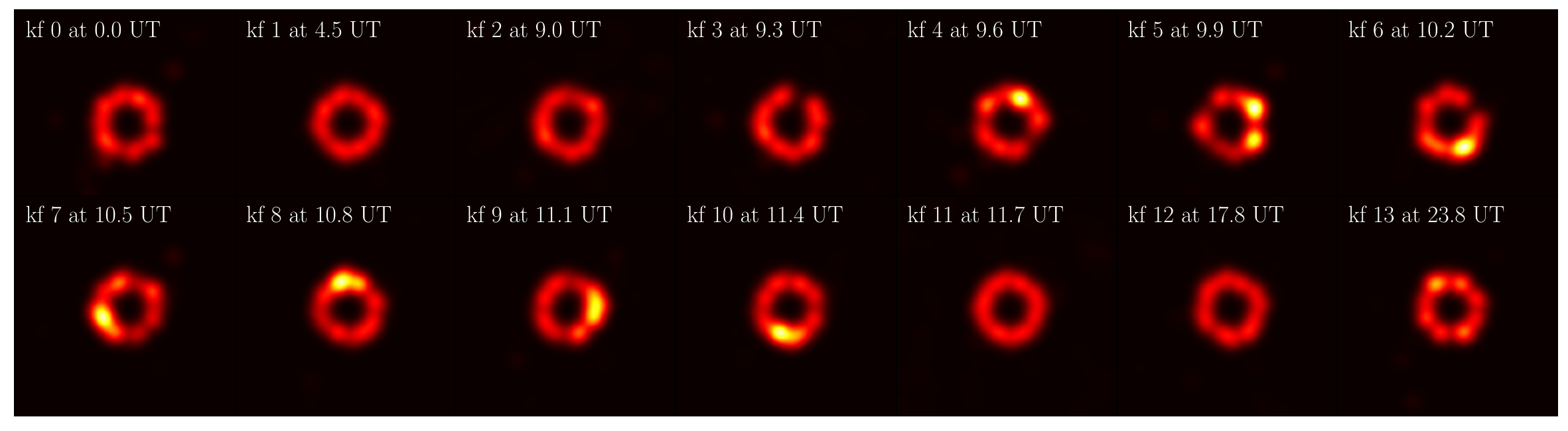}
  \caption{\sgra plus a transient hotspot, 14 keyframes}
  \label{fig: disk_eht}
\end{subfigure}%
\begin{subfigure}{0.5\textwidth}
  \centering
  \includegraphics[width=0.4\linewidth]{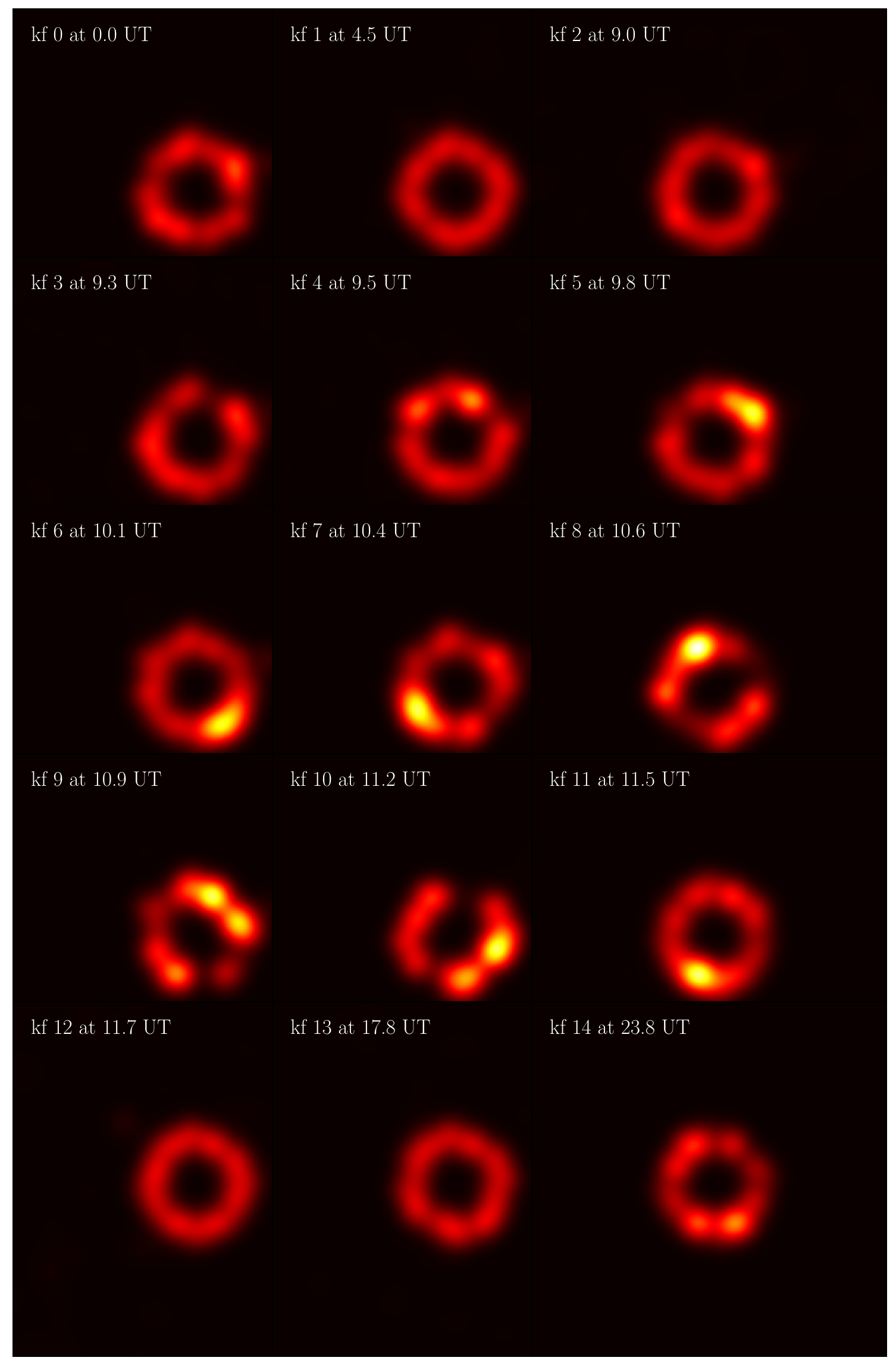}
  \caption{\sgra plus a transient hotspot, 15 keyframes}
\end{subfigure}

\begin{subfigure}{0.5\textwidth}
  \centering
  \includegraphics[width=0.8\linewidth]{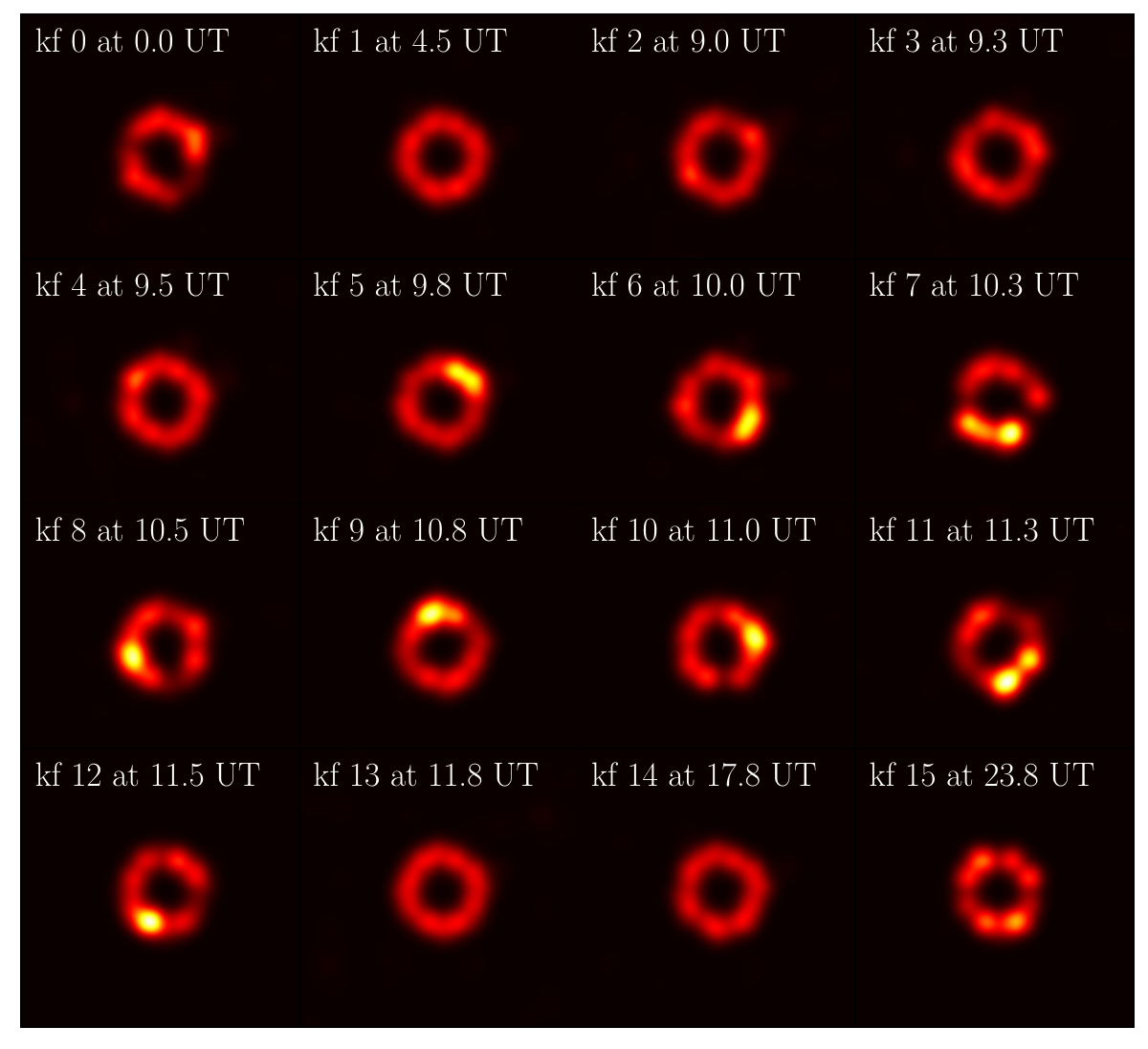}
  \caption{\sgra plus a transient hotspot, 16 keyframes}
  \label{fig: ring_eht}
\end{subfigure}%
\begin{subfigure}{0.5\textwidth}
  \centering
  \includegraphics[width=0.6\linewidth]{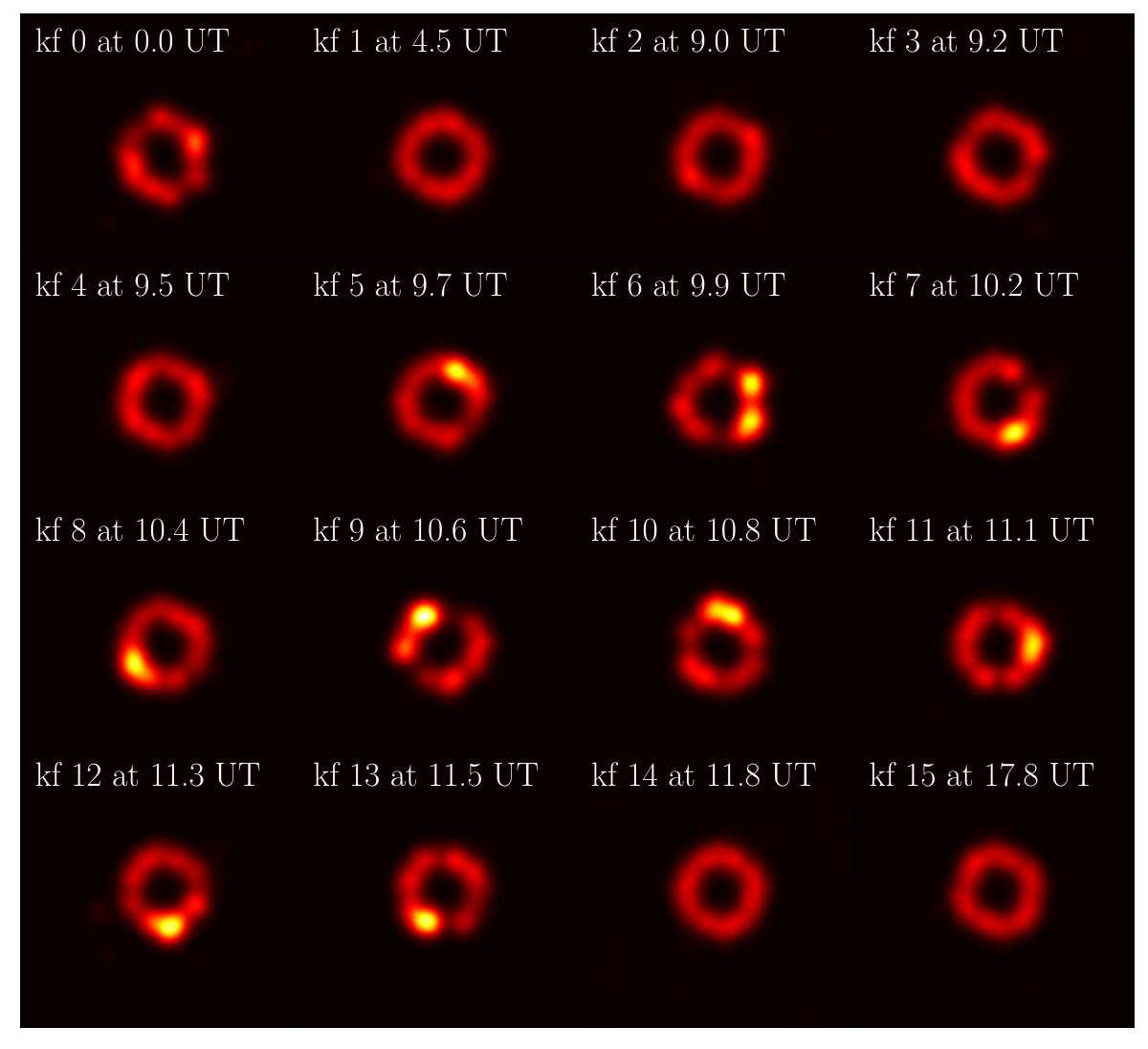}
  \caption{\sgra plus a transient hotspot, 17 keyframes}
\end{subfigure}

\caption{Solutions of the ngMEM for the case of \sgra plus a persistent hotspot (panels (a) and (b)) and \sgra plus a transient appearing during two hours of the observation (panels (c) to (f)). The field-of-view (fov) is 128 $\mu$as for all cases.}
\label{fig:eht_full}
\end{figure*}

\begin{figure*}
    \centering
    \includegraphics[width=\textwidth]{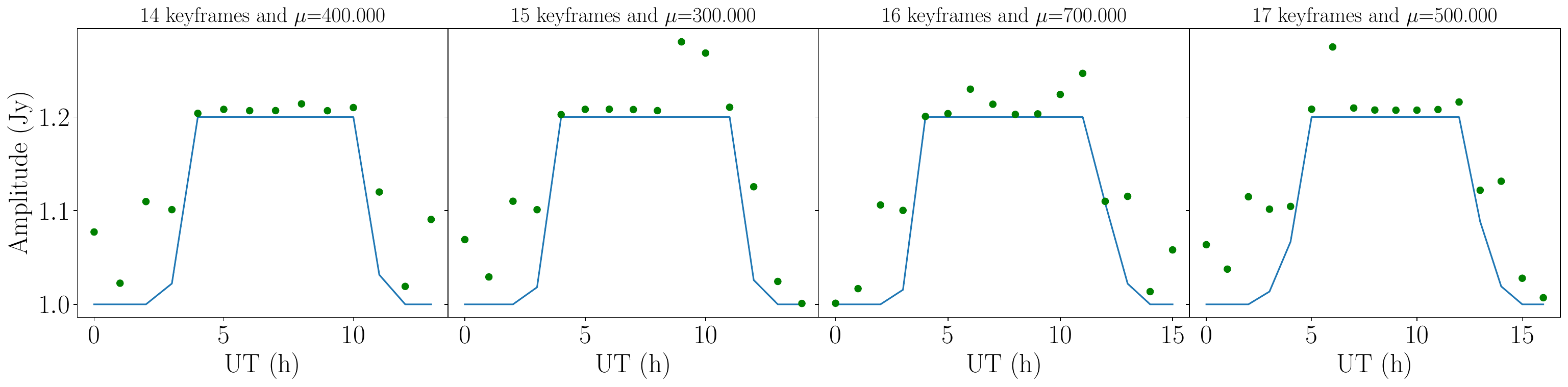}
    \caption{Groundtruth flux (continuous line) vs recovered flux (dots) by the ngMEM for the case of \sgra with an spurious hotspot. In continuous line, the flux's model, in bullets the values recovered by the algorithm. In the x-axis the UT and in the y-axis the amplitude (\,Jy).}
    \label{fig:my_label}
\end{figure*}

\begin{table*}
\caption{Free parameters}
\centering
\begin{tabular}{lrr}
\toprule
Number of keyframes &   $\mu$ &  $\tau$ \\
\midrule
10 (constant hotspot) &                900 &                  0.11 \\
14 (spurious hotspot) &                400 &                  0.08 \\
15 (spurious hotspot) &                300 &                  0.07 \\
16 (spurious hotspot) &                700 &                  0.07 \\
17 (constant hotspot) &                200 &                  0.06 \\
17 (spurious hotspot) &                500 &                  0.06 \\
\bottomrule
\label{tab:distance_15_simple}
\end{tabular}
\\
\textbf{Notes:} Values of the free-parameters: 1) number of keyframes, 2) time entropy regularizer $\mu$ and 3) image memory $\tau$ used for the movies presented in this Appendix.
\end{table*}

%% file: Appendix_initial_points.tex
\section{Comparison between different starting points}
\label{sec:comparison_initial_points}

In this appendix, we present a comparison of the results obtained when using different initial points for solving the ngMEM. As mentioned in Sect.~\ref{sec:ngmem_formalism}, non-convex problems may have multiple minima (maxima). Even though they are optimal solutions, and even if they satisfy KKT conditions, their associate value of the cost function may vary.\\
We present here an example of two sets of solutions of the ngMEM for the same problem: the hotspot orbiting during the 24 observation hours and 10 keyframes. The algorithm was run using two different initial points: the dirty image and the result of a first round of ngMEM with $\mu=0$.\\

\begin{figure*}
    \centering
    \includegraphics[scale=0.3]{PLOTS_CASE_TWO_DIM/SYNTHETIC_CASE/TIME_DEPENDENT/mfront_10.pdf}
    \includegraphics[scale=0.3]{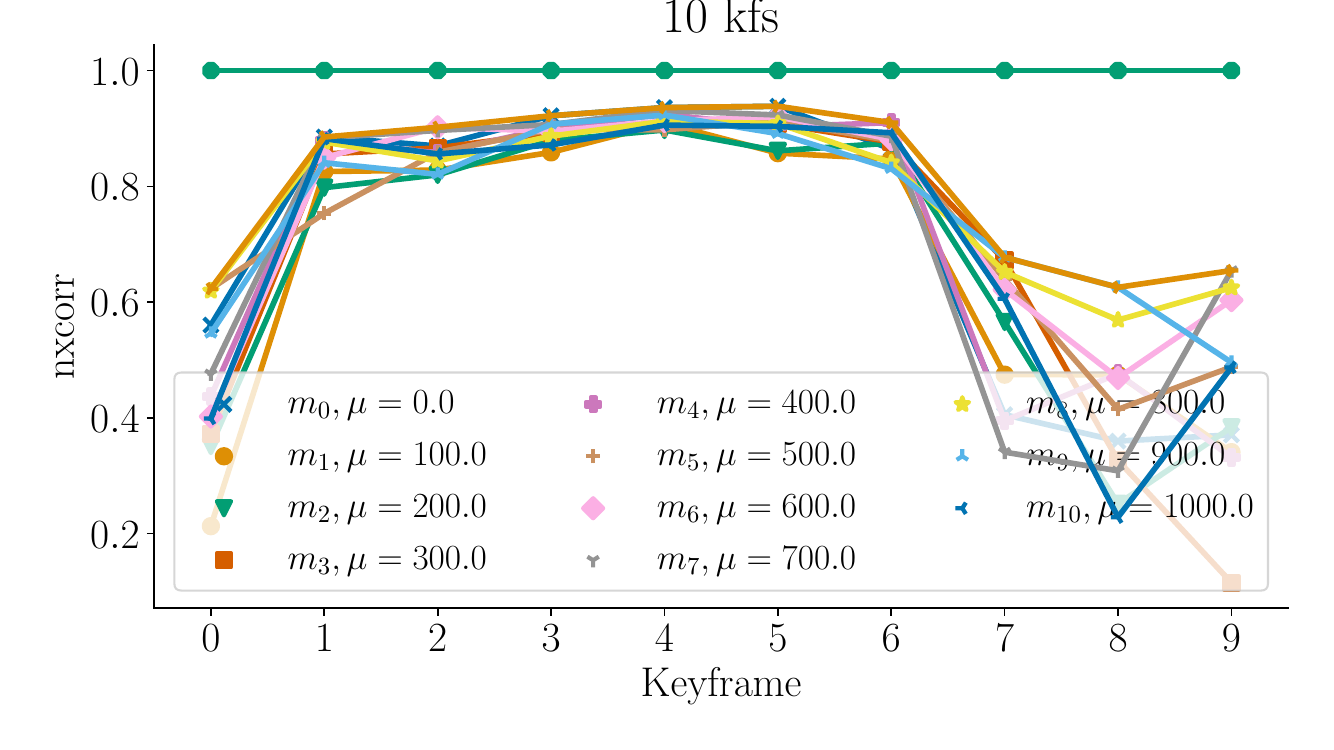}\\
    \includegraphics[scale=0.3]{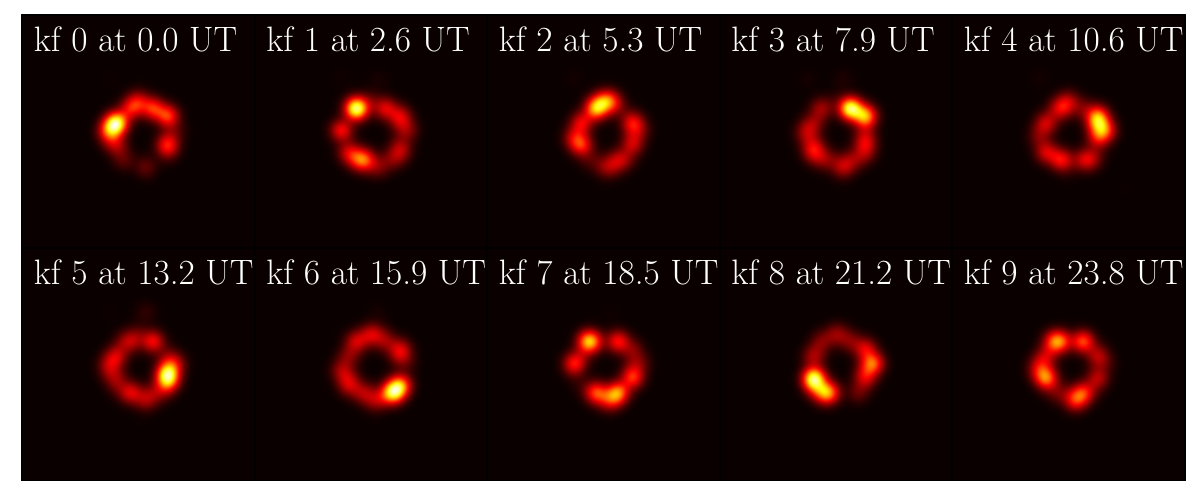}
    \includegraphics[scale=0.3]{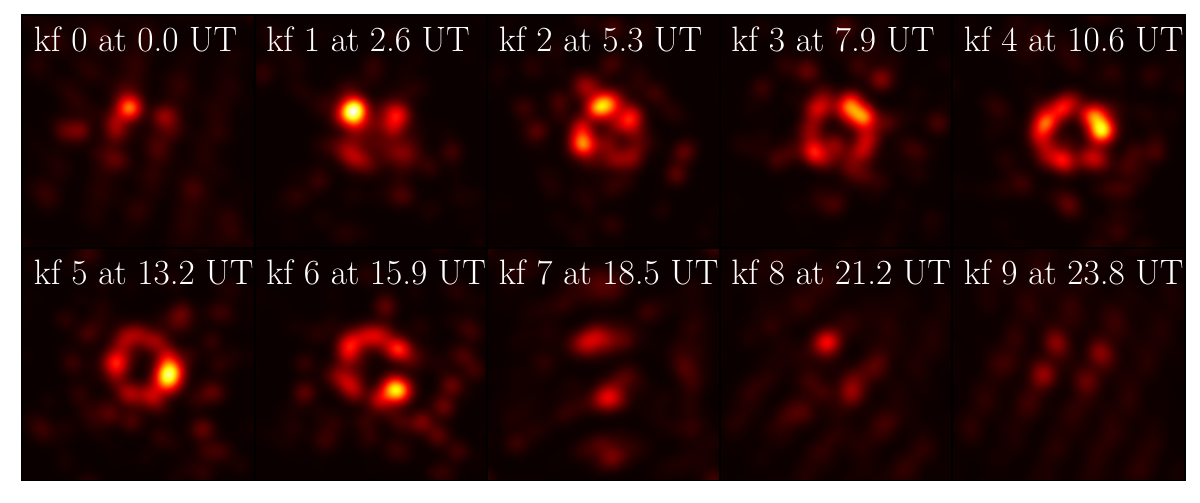}
    \caption{$\mfront$ comparison for the 24 hours orbiting hotspot using the dirty image (right column) and the result of the static ngMEM (left column) as the starting point of the algorithm.}
    \label{fig:comparison_initialpoints}
\end{figure*}

Figure~\ref{fig:comparison_initialpoints} shows the $\mfront$, the flux recovered and the movies obtained for their optimal $\mu$ (800 for the dirty image case and $\mu=900$ for the other one). Left column refers to the dirty image and right column to the one obtained using the static ngMEM ($\mu=0$).\\
The $\mfront$ already shows the low \nxcorr that starting with the dirty image has in some keyframes, even decreasing in some cases down to less than 0.2.\\
By comparing the movies with lesser $\mfront$ we can notice how the flux obtained is much worse in the first case, meaning that the equality constraint has been poorly satisfied in this local minimum. In overall, the movie produced starting with the dirty image, even if in most frames recovers the hotspot, is much noisier.

%% file: Appendix_static.tex
\section{Time-independent \MakeLowercase{ng}MEM}
\label{sec:time_dependent_ngmem}

In this appendix, we focus on the performance of the time-independent ngMEM and provide a comparison with the use of a RML method.

When reconstructing static images using ngMEM, it is sufficient to set $\mu=0$. Consequently, the super-Hessian described in Eq.~\eqref{eq: hessian_time_dep} is simplified to:

\begin{equation}
    \mathcal{H} = \begin{bmatrix}
    H^1 & \mathbf{0}    & \cdots & \mathbf{0} \\
    \mathbf{0} & H^2    & \cdots & \mathbf{0} \\
    \vdots & \vdots & \ddots & \vdots \\
    \mathbf{0} & \mathbf{0} & \cdots & H^{r} \\
    \end{bmatrix},
    \label{eq: hessian_time_indep}
\end{equation}

\begin{figure*}
    \centering
    \includegraphics[scale=0.05]{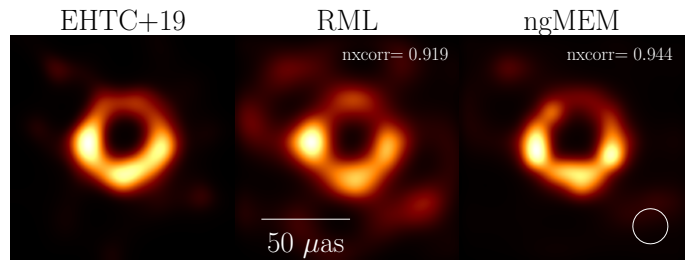}
    \vspace{-1cm}
    \caption{
    The reconstructed image of M87$^*$ using an RML method, as the one described in \citef{ehtimb}, is displayed in the center panel. On the right panel, the reconstruction obtained using the ngMEM with $\mu=0$ is shown. Both reconstructed images have been convolved with the same 20\,$\mu$as beam, represented by a white circle. The fov is indicated in the center image.
    To assess the similarity between the reconstructed images and the publicly available EHT image (shown in the left column), the \nxcorr values are displayed in the top corner of each image.}
    \label{fig:rml_ngmem}
\end{figure*}

Using \texttt{DoG-HIT}\footnote{software available at https://github.com/hmuellergoe/mrbeam}~\citep{Mueller22a, Mueller23a}, we have implemented Prob.~\eqref{eq:dual_mem_timeindp_other_version} with active regularizers for the $\chi^2$ for the visibilities, the entropy and total flux~\citep[see, for instance][]{ehtimb}. Then, we have imaged the SMBH Messier87$^*$, M87$^*$, observed with the uv-coverage from April 11, 2017~(\citealt{EHTM871} to~\citealt{EHTM876}). The results are depicted in Fig.~\ref{fig:rml_ngmem}. The left panel showcases the M87$^*$ image released by the EHTC to the public. In the center and right panels, we present the reconstructions obtained using the unconstrained problem and the static ngMEM, respectively. Both reconstructed images have been convolved with a 20\,$\mu$as beam.\
\\
